# Signal Improvement and Contrast Enhancement in Magnetic Resonance Imaging

by

Yi Han

Department of Chemistry
Duke University

Date: ________________________
Approved:

______________________________
Warren S. Warren, Supervisor

______________________________
Michael J. Therien

______________________________
Mark W. Dewhirst

# Abstract

This thesis reports advances in magnetic resonance imaging (MRI), with the ultimate goal of improving signal and contrast in biomedical applications. More specifically, novel MRI pulse sequences have been designed to characterize microstructure, enhance signal and contrast in tissue, and image functional processes. In this thesis, rat brain and red bone marrow images are acquired using iMQCs (intermolecular multiple quantum coherences) between intermediate separated spins. As an important application, iMQCs images in different directions can be used for anisotropy mapping and tissue microstructure analysis. At the same time, the simulations prove that the dipolar field from the overall shape only has small contributions to the experimental iMQC signal. Besides magnitude of iMQCs, phase of iMQCs should be studied as well. The phase anisotropy maps built by our method can clearly show susceptibility information in rat brain. It may provide meaningful diagnostic information. To deeply study susceptibility, the modified-crazed sequence is developed. Combining phase data of modified-crazed images and phase data of iMQCs images is very promising to construct microstructure maps. Obviously, the phase image in all above techniques needs to be highly-contrasted and clear. To achieve the goal, algorithm tools from Susceptibility-Weighted Imaging (SWI) and Susceptibility Tensor Imaging (STI) stands out superb useful and creative in our system.

# Contents

# List of Abbreviations

iMQCs – Intermolecular multiple quantum coherences

iDQCs – Intermolecular double quantum coherences

iZQCs – Intermolecular zero quantum coherences

CRAZED – COSY Revamped by Asymmetric Z-Gradient Echo Detection

DTI – Diffusion Tensor Imaging

SWI – Susceptibility Weighted Imaging

STI – Susceptibility Tensor Imaging

$\omega_0$ – nuclear Larmor frequency

$T_1$ – Longitudinal relaxation rate

$T_2$ – Transverse relaxation rate

TE – Echo time

$\sigma$ – Chemical shift

$\rho$ – Density matrix

$\gamma$ – Gyromagnetic ratio (for protons, this is 42.8 MHz/T or 2.68 x 108 rad/s/Tesla)

$k$ – Boltzmann constant (1.38 × 10-23 m2 kg s-2 K-1)

$\hbar$ - Planck's constant divided by $2\pi$ (6.626 x 10-34 J*s)

$\rho_0$ - Spin density

# List of Figures

# 1. Introduction to Magnetic Resonance Imaging

This thesis provides a description of new developments in signal improvement and contrast enhancement for magnetic resonance imaging. It is easy to understand how these new techniques lead to novel developments if we deeply understand the classical and quantum mechanical description of magnetic resonance. Thus, a brief overview of the most critical aspects of magnetic resonance including both the classical and quantum perspectives will be covered in the first chapter. More detailed descriptions can be found in Refs. [2-4].

## 1.1 Classical description of magnetic resonance

### 1.1.1 Spin and related basic concepts

Nuclear spin is the basis for all magnetic resonance research. It is a form of angular momentum that is not produced by a rotation of the particle, but rather is an intrinsic property of the particle itself. Spin behaves as expected from a classical description of a rotating object and fits the quantum mechanical descriptions of angular momentum. The total angular momentum squared of a particle with spin I takes values of the form [I(I+1)]1/2, with 2I+1 possible sublevels. The spin quantum number, I, depends on the nature of the particle – for fermions, this value is half integer; for bosons, this number is an integer. The applications described here deal with systems that are spin 1/2.

The slight excess of spins aligned in one direction yields a net magnetic moment, which is the signal detected in an NMR experiment. When a magnetic moment is placed in a magnetic field, it will tend to align with the field via the Zeeman interaction. A magnetic moment can be thought of as a bar magnet and the influence toward alignment can be described as the torque on the bar magnet exerted by the magnetic field. According to electromagnetic theory [5], the Hamiltonian (quantum mechanical operator for energy) of a magnetic moment $\vec{\mu}$ placed in a field $\vec{B}$ is $H = -\vec{\mu} \cdot \vec{B}$ and the torque is $\vec{\mu} \times \vec{B}$. The latter formula means that the rate at which the spin precesses depends on the magnetic field as well as the characteristics of the spin. Initially, the isotropically-oriented spins precess on the order of nanoseconds about the field, where the precession frequency is given by:

$$\omega = -\gamma B \tag{0.1}$$

where $B$ is the magnetic field at the site of the particle and $\gamma$ is the gyromagnetic ratio (2.68 x 108 rad/s/Tesla for 1H). For nuclear spins, $\omega$ is called the nuclear Larmor frequency. The Larmor frequency is proportional to the magnetic field.

### 1.1.2 Bloch equation

Rather than consider a single spin, Bloch [6] showed that it was possible to consider the ensemble of spins. More precisely, the net magnetization for an ensemble of spin 1/2 nuclei is:

$$M_0 = \frac{\rho_0 \gamma^2 \hbar^2 B_0}{4kT} \tag{0.2}$$

The spin density is $\rho_0$, $k$ is the Boltzmann constant, 1.380 x $10^{-23}$ J·K$^{-1}$, $\hbar$ is Planck's constant divided by $2\pi$, 1.054 x $10^{-34}$ J·s, and $T$ is the temperature in Kelvin.

The above describes the net magnetization. More generally, the total magnetic moment (magnetization) of a sample can be described as $M = \sum_i \mu_i$. In addition to consider that $\vec{\mu} = I \cdot A = \frac{qv}{2\pi r} \times \pi r^2 = \frac{q}{2m} \vec{L} = \gamma \vec{L}$, the rate of change in angular momentum gives the fundamental equation of motion for magnetization:

$$\frac{d\vec{M}}{dt} = \gamma \vec{M} \times \vec{B} \tag{0.3}$$

For example, consider a static magnetic field $B_o$ applied along $\hat{z}$. This is the simplest case in the free precession situation in the absence of relaxation and RF. The solutions to these are:

$$\begin{aligned} M_x &= M_{x0} \cdot \cos(\omega t) + M_{y0} \cdot \sin(\omega t) \\ M_y &= M_{y0} \cdot \cos(\omega t) - M_{x0} \cdot \sin(\omega t) \end{aligned} \tag{0.4}$$

where $\omega = \gamma B_o$ The results above account only for precession when in fact the magnetization vectors undergo relaxation processes which alter their amplitude and direction. In 1946 Felix Bloch [6] proposed a simple model to describe relaxation. It turns out that the $\hat{z}$ component of magnetization in a static field exhibits a trend toward its equilibrium value $M_0 : dM_z / dt = -\left(M_z - M_0\right) / T_1$ with $T_1$ as the time constant, also known as the longitudinal relaxation time. It indicates the time it takes for $M_z$ to be restored to its equilibrium value of $M_0$. T1 relaxation is the result of mostly non-radiative interactions between the spin system and the lattice (which connects the spin system and the external world). Relative to other types of spectroscopy, T1 is fairly slow (usually on the order of seconds), reflecting the weak interactions between the spins and their environment.

The transverse ( $\hat{x}$ and $\hat{y}$ ) components, on the other hand, are characterized by a decay due to interaction of spins with rapidly fluctuating magnetic fields from their surroundings which can be described as $dM_1 / dt = -M_1 / T_2$ and $dM_2 / dt = -M_2 / T_2$. $T_2$ is known as the transverse relaxation time. $T_2$ relaxation comes from the spin-spin interactions, which cause the Larmor frequency of the spins to change and dephasing the magnetization. In the absence of broadening effects (such as those from magnetic field imperfections), the spin-spin interactions determine the width of the peak, where the peak shape is Lorentzian (due to the Fourier transform of an exponential decay).

In 1956, Torrey [7] introduced an additional term $D\nabla^2\vec{M}$ to explain the extra damping of the magnetization due to self-diffusion. The resulting equations of motion are:

$$\frac{d\vec{M}}{dt} = \gamma \vec{M} \times \vec{B} - \frac{M_x \hat{x} + M_y \hat{y}}{T_2} - \frac{M_z - M_0}{T_1} \hat{z} + D\nabla^2 \vec{M} \qquad (0.5)$$

### 1.1.3 Vector diagram representation

Many experiments can be explained correctly using a vector diagram representation, where the behavior of the system under different conditions can be drawn and understood simply. An example, spin echo sequence is shown in figure 1. In magnetic resonance, a spin echo is the refocusing of spin magnetization by a pulse of resonant electromagnetic radiation. It is frequently used in NMR and MRI.

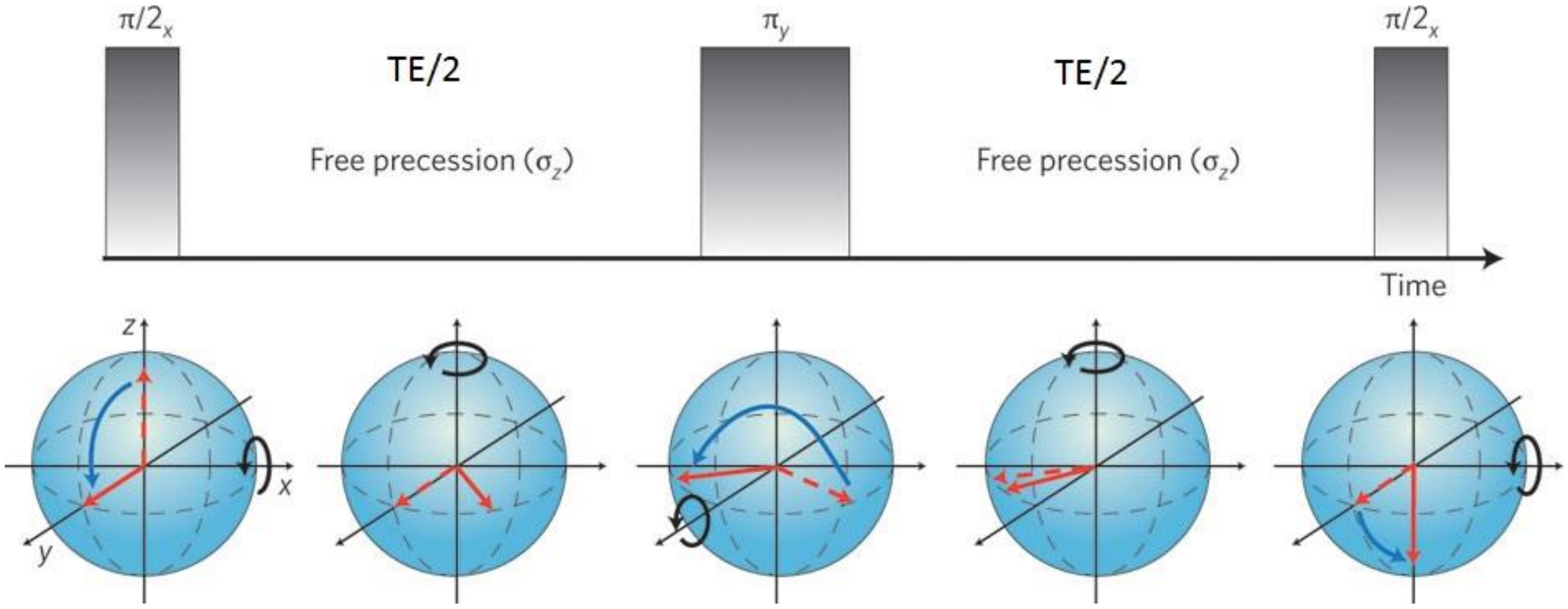

**Figure 1: Vector diagram of a spin echo [1]. On the left, a spin (dashed red arrow) is rotated to the x-y plane with a π/2 pulse. Magnetic fields then cause the spin to precess anticlockwise around the z-axis at Larmor frequency rate. By rotating the spin 180 degree around the y-axis in the middle of the free precession period with a π pulse, the original spin orientation is recovered, allowing another spin manipulation to occur.**

In this sequence, the net magnetization is initially oriented along the net magnetic field which is the z-direction. A 90-x RF pulse is applied to the sample, which rotates the magnetization vector into the x-y plane. The magnetization is allowed to precess, where the vector rotates around the net magnetic field at the Larmor frequency and becomes dephased due to T2. The time that spins freely precess is referred to as the echo time TE. TE is usually defined as the time between the 90 degree pulse and the center of the echo. After a time TE/2, 180 degree RF pulse is given, which reverses the evolution and causes refocusing of the signal TE/2 later [1].

Although the Bloch equations and vector diagram representation stated above can accurately describe many experiments, more complex phenomena such iMQCs are better understood with a quantum mechanical description which will be discussed later.

## 1.2 Quantum description of magnetic resonance

We review the basic principles of quantum mechanics relevant to NMR in this section, in order to better understand more complicated phenomena, such as the interactions of coupled spins or multiple quantum experiment. Hamiltonian operator $H$ is the most important object which determines the energy spectrum of a system and governs its time evolution through the Schrodinger equation. First of all, we need to know the basics of NMR states in order to know Hamiltonian.

### 1.2.1 NMR states

In quantum mechanics, a spin 1/2 nucleus has two possible orientations when placed in a magnetic field, spin up and spin down. We call these two stationary states as “eigenstates”, which are the possible energy states for the spins and by no means the only states available to the system. It means a spin can occupy these states statically for eternity, making these the states which are occupied at equilibrium. In magnetic resonance, these eigenstates are described as $|\alpha\rangle$ (aligned with the magnetic field, lower energy level) and $|\beta\rangle$ (aligned against the magnetic field, higher energy level) [3, 6, 8]. These two states form an orthonormal basis. In general, with light, we can temporarily put the spin into any state which is a linear combination of these eigenstates. For example, in a system of isolated spin 1/2 nuclei, the wavefunction can be written as:

$$\psi(t) = c_{\alpha}(t)|a\rangle + c_{\beta}(t)|\beta\rangle \tag{0.6}$$

where the two coefficients above are related to the populations in each state.

For multiple spins, since each spin has two different states, there are more states; for example, two spins have four possible configurations $|\alpha\alpha\rangle$, $|\alpha\beta\rangle$, $|\beta\alpha\rangle$ and $|\beta\beta\rangle$. Note that the order of the states indicates which spin is in the $|\alpha\rangle$ or $|\beta\rangle$ state. They might also have some coefficients which indicate possibilities being in that state.

Although these coefficients are often complex numbers, we are observing the eigenvalues of Hermitian matrices, which are always real-valued. Rather, the complex notation is used because there are two components of the magnetization that are 90° out of phase with each other, and it is convenient to represent these vectors using the complex notation.

As mentioned in the previous section, the observable signal in magnetic resonance comes from the ensemble of spins, and not from one isolated spin. The Hamiltonian describes all interactions of the entire system, and the wavefunction contains all positions, velocities, interactions and spin states of every nuclei and electron in the sample. While the Hamiltonian and wavefunction are the most complete descriptions of the system, they are not solvable in most realistic situations. Instead, it is commonly assumed that the electrical and magnetic influences of the rapidly moving electrons are blurred to the extent that only their average contribution can be seen, and thus, the Hamiltonian can be reduced to simply the nuclear spin Hamiltonian.

### 1.2.2 The Hamiltonian

As mentioned above, the Hamiltonian for a magnetic moment in a magnetic field is $H = -\vec{\mu} \cdot \vec{B}$. There is a slight tendency to align in the lower energy level, with exponential growth of the polarization over T1 period. The eigenstates of the Hamiltonian are the energy levels, which depend on the gyromagnetic ratio:

$$E_i = m_i \hbar \omega_0 = m_i \hbar \gamma B_0 \tag{0.7}$$

where $m_i$ is the magnetic quantum number for the nuclear spin. The separation between these two energy levels increases with increasing $B_0$. The selection rules allow only transitions separated by $\Delta m = \pm 1$.

If we consider external spin interactions in the static magnetic field and those in the oscillating transverse field:

$$\begin{aligned} H_{static} &= -\gamma B_0 S_z \\ H_{RF} &= -\frac{1}{2}\gamma B_{RF}\left(\cos\left(\omega_1 T\right) I_x + \sin\left(\omega_1 T\right) I_y\right) \end{aligned} \tag{0.8}$$

where the radio frequency interaction is only present during the RF pulses. In addition to these external spin interactions, there are also internal spin interactions. It includes chemical shift, J-coupling, dipole-dipole couplings, quadrupolar couplings, and spin-rotation interactions. Chemical shift and J-coupling have important effects determining the structural determination of organic molecules. In iMQCs the dipolar couplings play a critical role.

Chemical shift terms arise from the indirect interaction between $B_0$ and nuclear spins that is mediated through electrons. Several factors include electron density, electronegativity of neighboring groups and anisotropic induced magnetic field effects. J-coupling is an indirect interaction between two nuclear spins which comes from hyperfine interactions between the nuclei and local electrons. Dipolar coupling results from each spin generating a magnetic field that is oriented parallel to the nuclear spin vector.

### 1.2.3 J coupling

Coupled spins means two or more spins are interacted with each other. J-coupling, the interactions between two nuclear spins through electrons in chemical bonds, is the most profound coupling effects in an isotropic liquid. The full form of J-coupling between spins $I_i$ and $I_j$ on the same molecule is:

$$H = 2\pi J_{ij}\hat{I}_i \cdot \hat{I}_j = 2\pi J_{ij}\left(\hat{I}_{iz} \cdot \hat{I}_{jz} + \hat{I}_{iy} \cdot \hat{I}_{jy} + \hat{I}_{ix} \cdot \hat{I}_{jx}\right) \tag{0.9}$$

where $\hat{I}_{iz} \cdot \hat{I}_{jz}$, $\hat{I}_{iy} \cdot \hat{I}_{jy}$ and $\hat{I}_{ix} \cdot \hat{I}_{jx}$ are so-called product operator here. $J_{ij}$ is the proportionality constant of the coupling (also referred to as the spin-spin coupling constant or J-value).

In summary, combining the relevant external and internal spin interactions, the Hamiltonian is:

$$\begin{aligned} H &= \sum_i \omega_0 I_z + \sum_{i<j} 2\pi J_{ij} I_i \cdot I_j \\ \omega_0 &= \gamma B_0 \left(1 - \sigma_i\right) \end{aligned} \tag{0.10}$$

where $\sigma_i$ is the chemical shift of spin i.

In the case of a weakly coupled system, which means the coupling constant of the spin pair is much weaker than the difference in chemical shift, the transverse component is ignorable so the J coupling Hamiltonian can be written as:

$$H = 2\pi J_{ij} \hat{I}_{iz} \cdot \hat{I}_{jz} \tag{0.11}$$

This is what we call the secular approximation which is very advantageous.

In many cases such as free procession, we always need to use this term $e^{-iHt}$. The Hamiltonian in this term can be made up of many components, such as chemical shift evolution and J coupling terms. As we know, $I_z$, $I_y$ and $I_x$ don't commute to each other, while they commute to themselves. Thus, if we drop all non-$I_z$ terms or the secular approximation, generally this term can be broken up as follows, $e^{A+B} = e^A e^B$. In NMR, it really makes our computation performable and easy.

The effect of the secular approximation is just that all non-$I_z$ terms of the Hamiltonian are ignored. In terms of the matrix representation, this is because the diagonal terms are exactly the $I_z$ terms, and all of the $I_x$ and $I_y$ terms are off-diagonal. In the secular approximation, the diagonal terms dominate when the chemical shift difference is much greater than the J coupling constant (in other words, in the weak coupling limit). Under this condition, everything but the $I_z$ terms can be dropped, and all of the remaining terms commute. Thus, the Hamiltonian exponential can be broken into the product of each individual $I_z$ -term and evaluated individually. Outside the weak coupling limit, the full expression for J coupling is necessary.

### 1.2.4 The density matrix

When we are going to understand the behavior of the ensemble, it would be necessary to sum over all the spins in the system, thus making for a difficult calculation. The density matrix formalisms allow us to understand the behavior of the system as a whole and predict the outcome of an NMR experiment.

If we consider an ensemble of spins, to a good approximation each spin behaves like an isolated spin [2]. Once this ensemble is placed in a magnetic field, some proportion (given by the Boltzmann distribution) of the spins will assume the $|\alpha\rangle$ state, and others will take the slightly higher energy $|\beta\rangle$ state. The net magnetization of the system is a sum of all of the tiny contributions from each spin, but the actual exact calculation of the net magnetization (by considering the contributions from every individual spin) is impractical because of the size of the ensemble.

The density operator formalism stands out as a practical method. It considers the quantum state of the whole ensemble instead of each spin state. For an arbitrary superposition state $|\psi\rangle = c_\alpha |a\rangle + c_\beta |\beta\rangle$, the expectation value of a given operator $\hat{P}$ can be written as $\langle P\rangle = \langle \psi | \hat{P} | \psi \rangle$. It is mathematically equivalent to $Tr\left(|\psi\rangle\langle\psi|\hat{P}\right)$, where $Tr$ represents the trace of a matrix and $|\psi\rangle\langle\psi|$ is the projection operator. So obviously below two equations have the same meanings:

$$\langle P\rangle = \langle\psi_1|\hat{P}|\psi_1\rangle + \langle\psi_2|\hat{P}|\psi_2\rangle + \langle\psi_3|\hat{P}|\psi_3\rangle + ... \tag{0.12}$$

$$\langle P\rangle = Tr\left(|\psi_1\rangle\langle\psi_1| + |\psi_2\rangle\langle\psi_2| + |\psi_3\rangle\langle\psi_3| + ...\right)\hat{P} \tag{0.13}$$

It tells us that instead of doing calculations for each individual state, equation (1.13) makes it possible to use the sum across states to do the calculation. This is the core concept of density operator formalism.

We can write the matrix form of the projection operator, it also gives the definition of density operator $\hat{\rho}$

$$\hat{\rho} = \overline{|\psi\rangle\langle\psi|} = \overline{\begin{pmatrix} c_\alpha \\ c_\beta \end{pmatrix}\begin{pmatrix} c_\alpha^* & c_\beta^* \end{pmatrix}} = \begin{pmatrix} \overline{c_\alpha c_\alpha^*} & \overline{c_\alpha c_\beta^*} \\ \overline{c_\beta c_\alpha^*} & \overline{c_\beta c_\beta^*} \end{pmatrix} = \begin{pmatrix} \rho_{\alpha\alpha} & \rho_{\alpha\beta} \\ \rho_{\beta\alpha} & \rho_{\beta\beta} \end{pmatrix} \tag{0.14}$$

where the bar denotes an ensemble average over all spins. Thus, the preceding density matrix can also be re-written as an expression of these other operators:

$$\begin{aligned} \hat{\rho} &= \rho_{\alpha\alpha}\begin{pmatrix} 1 & 0 \\ 0 & 0 \end{pmatrix} + \rho_{\alpha\beta}\begin{pmatrix} 0 & 1 \\ 0 & 0 \end{pmatrix} + \rho_{\beta\alpha}\begin{pmatrix} 0 & 0 \\ 1 & 0 \end{pmatrix} + \rho_{\beta\beta}\begin{pmatrix} 0 & 0 \\ 0 & 1 \end{pmatrix} \\ &= \rho_{\alpha\alpha}\hat{I}_\alpha + \rho_{\alpha\beta}\hat{I}_+ + \rho_{\beta\alpha}\hat{I}_- + \rho_{\beta\beta}\hat{I}_\beta \end{aligned} \tag{0.15}$$

This expression clearly shows us the advantages of the density matrix and its statistical properties of populations and coherences. The diagonal elements, $\rho_{\alpha\alpha}$ and $\rho_{\beta\beta}$, represents the statistical population of spins in the spin-up $\hat{I}_\alpha$ and spin-down $\hat{I}_\beta$ states. The off-diagonal elements, $\rho_{\alpha\beta}$ and $\rho_{\beta\alpha}$, are the coherences between those two states. It tells us, using statistical mechanics, the correlations of the phase changes from "down-to-up" $\hat{I}_+$ and from "up-to-down" $\hat{I}_-$, respectively. Since the $\hat{I}_+$ represents the transition from the state with higher energy to the state with lower energy, $|\beta\rangle$ to $|\alpha\rangle$, its coefficient $\rho_{\alpha\beta}$ corresponds to a detectable NMR signal.

In general at thermal equilibrium, the density matrix will have no coherence (off-diagonal elements are zero) and its populations will obey the Boltzmann distribution:

$$\hat{\rho}_{eq} = \frac{e^{\frac{H}{k_B T}}}{Tr\left(e^{\frac{H}{k_B T}}\right)} = \frac{e^{\sum_i \frac{-\hbar\omega_0}{k_B T}\hat{I}_{zi}}}{Tr\left(e^{\sum_i \frac{-\hbar\omega_0}{k_B T}\hat{I}_{zi}}\right)} = \frac{\prod_i e^{\frac{-\hbar\omega_0}{k_B T}\hat{I}_{zi}}}{Tr\left(\prod_i e^{\frac{-\hbar\omega_0}{k_B T}\hat{I}_{zi}}\right)} \tag{0.16}$$

where $k_B = 1.38066\times10^{-23} J\cdot K^{-1}$ is the Boltzmann constant. We can re-write the exponential by hyperbolic functions:

$$\hat{\rho}_{eq} = \frac{\prod_i e^{-\frac{\hbar\omega_0}{k_B T}\hat{I}_{zi}}}{Tr\left(\prod_i e^{-\frac{\hbar\omega_0}{k_B T}\hat{I}_{zi}}\right)} = \frac{\prod_i \left(\cosh\left(-\frac{\hbar\omega_0}{2k_B T}\right)\hat{E} - 2\sinh\left(-\frac{\hbar\omega_0}{2k_B T}\right)\hat{I}_{zi}\right)}{Tr\left(\prod_i \left(\cosh\left(-\frac{\hbar\omega_0}{2k_B T}\right)\hat{E} - 2\sinh\left(-\frac{\hbar\omega_0}{2k_B T}\right)\hat{I}_{zi}\right)\right)}$$

$$= \frac{\prod_i \left(\cosh\left(-\frac{\hbar\omega_0}{2k_B T}\right)\hat{E} - 2\sinh\left(-\frac{\hbar\omega_0}{2k_B T}\right)\hat{I}_{zi}\right)}{2^N \cosh^N\left(-\frac{\hbar\omega_0}{2k_B T}\right)} \tag{0.17}$$

$$= 2^{-N}\prod_i \left(\hat{E} - 2\tanh\left(-\frac{\hbar\omega_0}{2k_B T}\right)\hat{I}_{zi}\right)$$

At room temperature, it goes further to:

$$\hat{\rho}_{eq} \approx 2^{-N}\left(\hat{E} + \left(-\frac{\hbar\omega_0}{k_B T}\right)\sum_i \hat{I}_{zi} + \frac{1}{2}\left(-\frac{\hbar\omega_0}{k_B T}\right)^2 \sum_i\sum_j \hat{I}_{zi}\hat{I}_{zj} + \ldots\right) \tag{0.18}$$

The high temperature approximation ignores all higher order terms after quadratic terms.

# 2. Introduction to Intermolecular Multiple Quantum Coherences

Intermolecular multiple quantum coherences (iMQCs) correspond to simultaneous transitions of nuclear spins on different molecules with a macroscopic separation, typically many microns [9-25]. The original (1948) framework of solution NMR precluded the existence of coherences between such widely separated (and weakly interacting) nuclei [9, 26]. However, a series of groundbreaking papers [16, 19, 25, 27] showed that theses coherences existed in solution and could be converted into strong observable signals with the proper pulse sequence. So a review of the conventional wisdom on dipolar effects in solution NMR will be introduced in this chapter. Next, how iMQCs was discovered, some corrections applied and how the corrections explain the appearance of iMQCs will be stated. Finally, having established how the dipolar coupling can reappear between distant spins, we analyze the pulse sequence which uses this to make signal from iMQCs experiments.

## 2.1 The traditional dipole interaction

In a material consisting of particles having dipole moments, each generating a dipole magnetic field, the net effect is the generation of a dipolar field which is the sum of all dipoles in the sample. The electromagnetic theory tells us that the magnetic field is the curl of the vector potential, $\vec{B}(\vec{r}) = \nabla \times \vec{A}(\vec{r})$. The magnetic vector potential of a dipole moment $\vec{m}$ is given by $\vec{A}(\vec{r}) = \frac{\vec{m} \times \hat{r}}{r^2}$. So:

$$\begin{aligned}
\vec{B}(\vec{r}) &= \nabla \times \left( \vec{m} \times \frac{\hat{r}}{r^2} \right) = \left( \frac{\hat{r}}{r^2} \cdot \nabla \right) \vec{m} - (\nabla \cdot \vec{m}) \frac{\hat{r}}{r^2} - (\vec{m} \cdot \nabla) \frac{\hat{r}}{r^2} + \left( \nabla \cdot \frac{\hat{r}}{r^2} \right) \vec{m} \\
&= -(\vec{m} \cdot \nabla) \frac{\hat{r}}{r^2} + \left( \nabla \cdot \frac{\hat{r}}{r^2} \right) \vec{m} \\
&= \frac{1}{|r|^5} \left( 3\langle \vec{m}, \vec{r} \rangle \vec{r} - \vec{m} |r|^2 \right)
\end{aligned} \tag{0.19}$$

The transition to the continuum is done by replacing the individual dipole moments $\vec{m}$ by local magnetizations $\vec{M}(\vec{r}) d^3\vec{r}$ and integrating over all space to give the dipolar field.

$$\vec{B}(\vec{r}) = \iiint d^3\vec{r}\,' \frac{1}{|\vec{r} - \vec{r}\,'|^3} \left[ \vec{M}(\vec{r}\,') - \frac{3\langle \vec{M}(\vec{r}\,'), \vec{r} - \vec{r}\,' \rangle (\vec{r} - \vec{r}\,')}{|\vec{r} - \vec{r}\,'|^2} \right] \tag{0.20}$$

Because the Bloch equations are usually cast in the rotating frame, only the part which remains static is relevant to the spin dynamics. The following formula derives the invariant part of the dipolar field, which is also called "secular part" in the NMR literature. You can also understand the word "secular" here to be the observable dipolar terms under the Zeeman Hamiltonian. The "non-secular" parts in high field NMR are negligible.

$$\vec{B}(\vec{r}) = \iiint d^3\vec{r}\,' \frac{1-3\cos^2\theta_{rr'}}{2|\vec{r}-\vec{r}\,'|^3}\left[3\vec{M}_z(\vec{r}\,')\hat{z} - \vec{M}(\vec{r}\,')\right] \tag{0.21}$$

where $\cos\theta_{rr'} = \langle \vec{r}-\vec{r}\,', \hat{z}\rangle / |\vec{r}-\vec{r}\,'|$. The dipolar interaction between two spins contributes positive or negative potential energy depending on the angle $\theta$ the vector adjoining to two nuclei makes with respect to the direction of the applied field. The angular factor $(3\cos^2\theta - 1)/2$ equals +1 for $\theta = 0$ or $\pi$, and $-1/2$ for $\theta = \pi/2$ .

Since the dipole interaction is inversely proportional to the cube of the separation between the two spins, it is generally considered as a local interaction (in the range of nanometers). The dipolar effect is significant in solid-state NMR because of the fixed relationship between spins. It results in severe line broadening and also splits peaks, even for magnetically equivalent spins. However, the situation is very different in solution. Since molecular tumbling occurs in solution, the angular dependency $1-3\cos^2\theta_{rr'}$ between any given intramolecular spin pair needs to be replaced by its time-averaged value $\langle 1-3\cos^2\theta_{rr'}\rangle$, which is equal to zero. So the two intermolecular spins interaction should also be averaged out by diffusion, which also occurs in solution.

Thus, the traditional NMR picture concludes that there is no observable dipolar effect in an isotropic liquid, although dipolar interactions do contribute to relaxation times. This conclusion has also been supported empirically: an ensemble of magnetically equivalent spins, like water, always shows a sharp singlet peak in solution NMR.

## 2.2 CRAZED experiments

Warren and co-workers proposed CRAZED, which is basically a COSY pulse sequence plus two gradients surrounding its second pulse, in 1993 [16, 25] (Figure 2.a). The ratio between the two gradients is well known as a "multiple quantum filter" used in NMR. A typical CRAZED can then be either zero quantum (ZQ) or double quantum (DQ), which solely depends on the ratio between the first and second gradient areas (the integral $\int_0^T G(\vec{r})\,dt$ of each gradient). If the ratio is 1:0, it is an iZQCs. If the ratio is 1:2, it is an iDQCs.

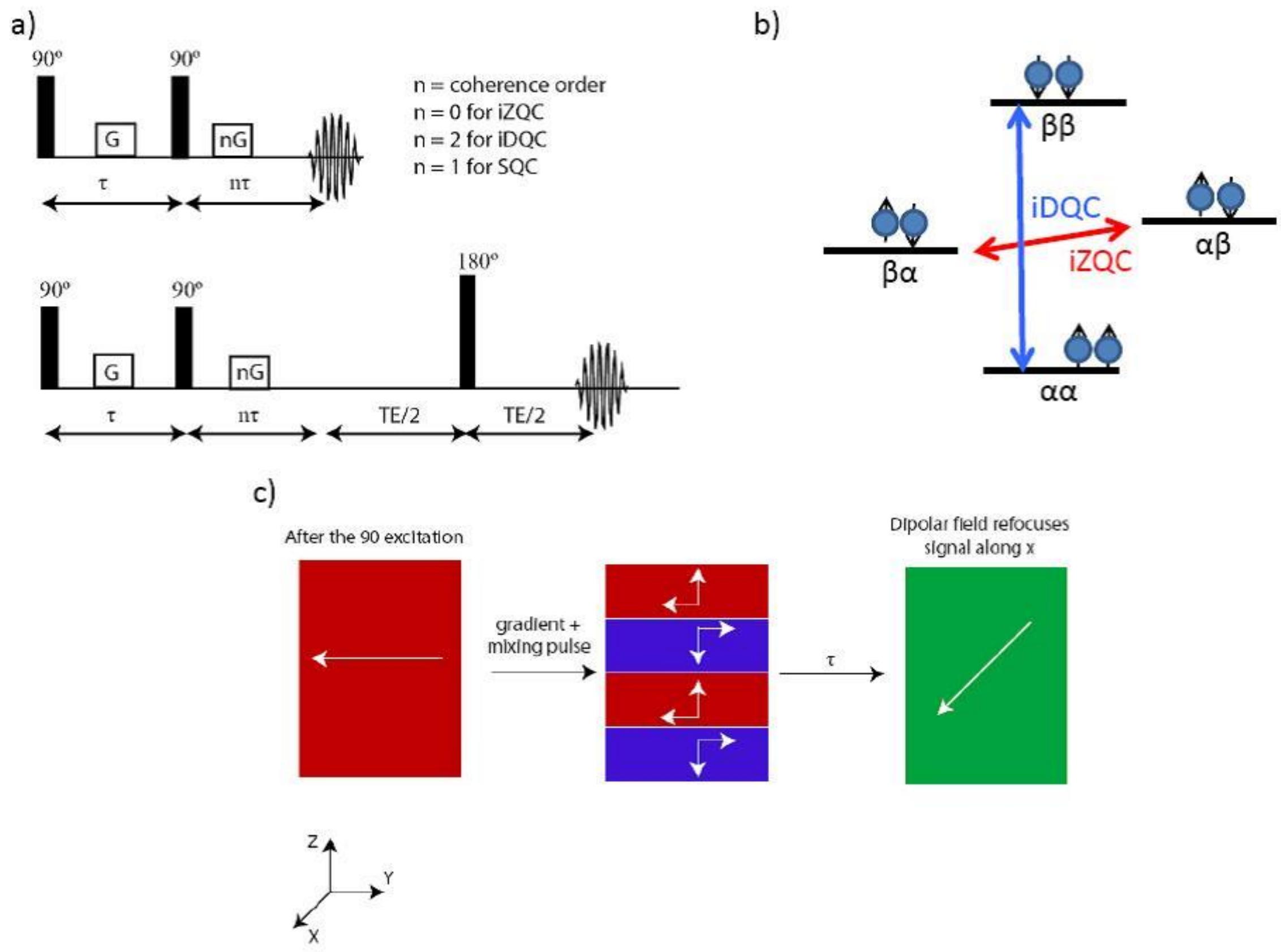

**Figure 2: (a). The CRAZED pulse sequence. The top sequence is the classic sequence, but it suffers from T2* weighting. The insertion of a refocusing pulse (below) reduces the effect of T2* relaxation. (b). Energy level diagram shows the DQC and ZQC transitions for a 2 spin system. (c). Pictoral description of the effects of the gradients and dipolar coupling. The gradients break the isotropy of the magnetic field, and reintroduce dipolar couplings. The dipolar couplings convert the unobservable 2-spin coherences into observable 1-spin coherences.**

Figure 3 is a typical iDQCs spectrum. The four peaks on this iDQCs spectrum are at $(-\Omega_w - \Omega_w, \Omega_w)$, $(-\Omega_A - \Omega_w, \Omega_w)$, $(-\Omega_w - \Omega_A, \Omega_A)$ and $(-\Omega_A - \Omega_A, \Omega_A)$, in which $\Omega_w$ and $\Omega_A$ are the offset frequencies of water and acetone, respectively. These peaks are double quantum peaks, whose frequencies in the F1 dimension are the combinations of their offset frequencies. However, these peaks were all forbidden in the traditional understanding of NMR if considering the pulse sequence in the figure 3.

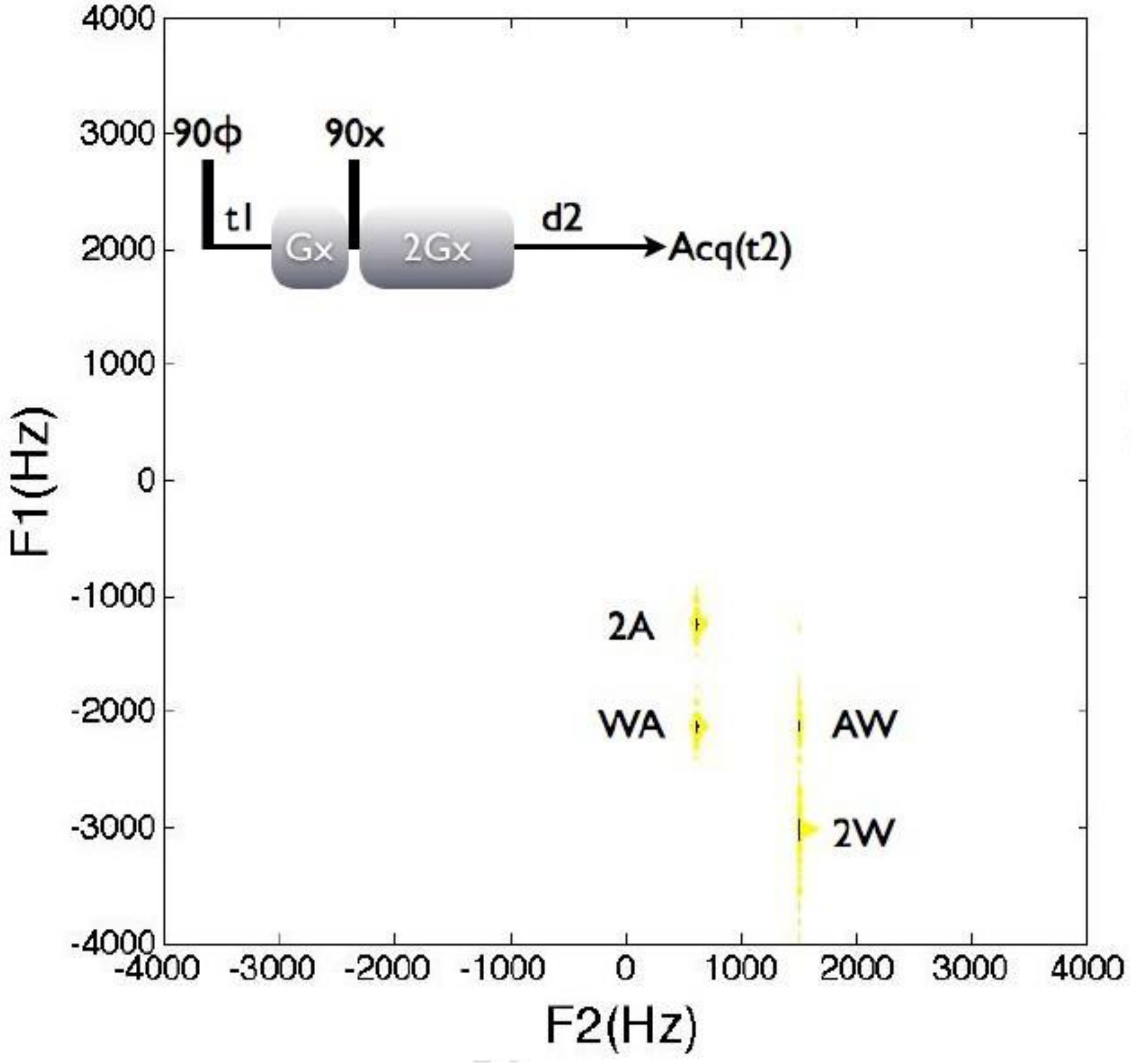

**Figure 3: Double Quantum COSY Spectrum of 1:1 Water/Acetone. A four step phase cycling ($\phi = x, y, -x, -y$) was performed to the first 90 degree pulse. The delay d2 was set to 60 ms. the water and acetone offsets were set at 1500 and 630 Hz, respectively. The 2W, AW, WA and 2A peaks are at (-3000, 1500), (-2130, 1500), (-2130, 630) and (-1260, 630), respectively.**

Detailed derivatives are in later sections. Now let's briefly look at this system. The six protons in acetone are magnetically equivalent to each other, like the two protons in water. The two homo-molecular double quantum peaks, $2W\left(-\Omega_w - \Omega_w, \Omega_w\right)$ and $2A\left(-\Omega_A - \Omega_A, \Omega_A\right)$, should not exist because: first, the J coupling should have no observable effect on these magnetically equivalent pairs; second, intramolecular dipolar interaction should have been averaged out by the molecular tumbling; and third, the intermolecular dipolar interaction between molecules that are close together should have been averaged out by diffusion. Following the same logic, the two hetero-molecular peaks, $AW\left(-\Omega_A - \Omega_w, \Omega_w\right)$ and $WA\left(-\Omega_A - \Omega_w, \Omega_A\right)$, should not exist, either. Furthermore, there are only two pulses in the CRAZED pulse sequence, which is one pulse less than the minimum requirement to create observable MQCs.

Since there was not supposed to be any effective coupling in this water/acetone system, it was previously treated as two ensembles of isolated spins. According to the vector model and Block equation previously, the magnetization would be dephased by first gradient and not refocused by the second gradient because of the 1:2 ratio. Thus, an experiment like iDQCs was not supposed to have any signal, but only noise.

The CRAZED experiment challenged people's understanding about spin dynamics. Obviously, some important things have been missed in these widely accepted pictures of NMR. In fact, the diffusion does not eliminate the dipolar interaction of every pair of spins. The high temperature approximation also oversimplifies the density matrix. Corrections are needed to both of these assumptions.

## 2.3 Correction one: distant dipolar field

There is no double that intramolecular spins interaction is averaged out by molecular tumbling. It is also true the diffusion eliminates the dipolar effects between nearby intermolecular pairs. In both cases, the inter-nuclear vector of each pair has sampled every $\theta_{rr'}$ in the NMR time scale so that the expectation value of $\langle 1-3\cos^2\theta_{rr'}\rangle$ is zero. The situation is different for molecules that are far apart: over the NMR time scale, diffusion cannot change their inter-nuclear vector very much, and the dipolar field is not averaged out. Diffusion will only average out intermolecular couplings inside a radius of $\sqrt{2Dt}$, where $D$ is the diffusion constant and $t$ is the given time scale. In a typical solution NMR experiment, the time scale is less than a second. For water, with a diffusion constant $D = 2.3\times10^{-9}\,m^2s^{-1}$ at room temperature, this radius is about 10 $\mu m$. For pairs separated much farther than 10 $\mu m$, each of these inter-nuclear vectors is relatively fixed so the diffusional average of $\langle 1-3\cos^2\theta_{rr'}\rangle$ does not end up zero.

Even though the distant coupling between two individual molecules is unimportant because the dipole coupling falls off as $|\vec{r}-\vec{r}\,'|^{-3}$, the sum of these distant couplings is not negligible. Considering the fact that the number of spins in the shell of distance $|\vec{r}|$ is proportional to $r^2$, not $r^3$, the field generated by these non-zero distant dipoles couplings, called the distant dipolar field (DDF), can be calculated by integrating over space. The $|\vec{r}|^{-1}$ distance dependency implies that the DDF generated by the spins between 10 ~ 500 $\mu m$ away is as big as the DDF generated by those only between 1 ~ 50 $nm$ away. However, the DDF still vanishes in an isotropic liquid unless the spherical symmetry is broken, since the spherical integral of $1-3\cos^2\theta_{rr'}$ is also zero for each shell when there is symmetry. A non-spherical sample can break the symmetry to reintroduce the distant dipolar field, even though this "shape-dependent DDF" is relatively small. The other way to recover the distant dipolar field is by creating spatially modulated magnetization, which makes each shell temporarily anisotropic.

## 2.4 Correction two: density matrix definition

In the previous section, a new method of density matrix is introduced. Let's recall the definition of the equilibrium density matrix ( $\rho_{eq}$ ). At thermal equilibrium, the populations of the energy states obey the Boltzmann distribution, as given by the equilibrium density matrix. The optimized expression for the density matrix of N identical spins under a Zeeman Hamiltonian is given in equation (1-18) at room temperature.

Since $\frac{\hbar\omega_0}{k_B T}$ is a very small number at room temperature, coefficients of $\left(\frac{\hbar\omega_0}{k_B T}\right)^2$ and $\left(\frac{\hbar\omega_0}{k_B T}\right)^3$ might be expected to render the later terms negligible. However, the third and higher order terms grow exponentially with the number of spins, N. Thus, these higher order terms should not simply be eliminated for an extra factor of Boltzmann constant, $10^{-4}$. It is from these higher-order terms in the density matrix that iMQCs derive their signal. It also can be seen later from sequences.

## 2.5 Derivations for CRAZED experiments

As you can see in figure 4, the degree of the second pulse (mixing pulse) is different from what's shown in figure 2.a. One reason is that different degrees leading to different coherences which will be discussed in this section. The other reason is that we also try to optimize out those degrees by searching the maximal signals.

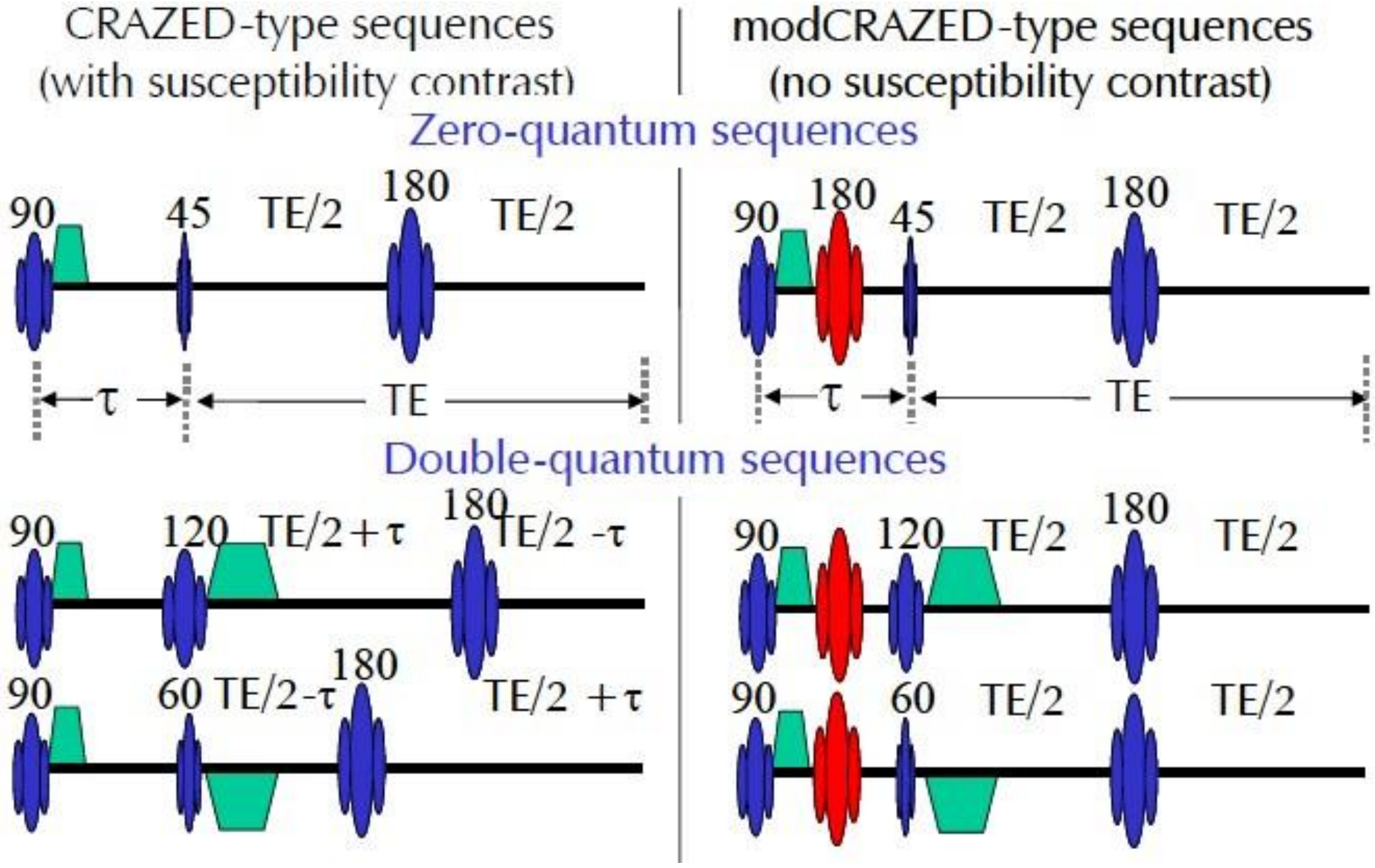

**Figure 4: CRAZED-type sequence (with susceptibility contrast) and modCRAZED-type sequences (no susceptibility contrast). In each type, zero-quantum sequences and double-quantum sequences are shown. In double-quantum sequences, +2 quantum double-quantum sequence and -2 quantum double-quantum sequence are shown.**

With above two corrections, two explanations (both correct) are put forward to describe the unexpected effects of iMQC signals. The first one is a quantum mechanical framework coming from density matrix and Hamiltonian. The second one is a classical description which uses revised block equation and the framework of the DDF. In this section, we are going to show you both explanations. Let's start from the quantum framework to explain iMQC signals.

Correction one says that in a liquid system the dipolar coupling can't be ignored if we break the spherical symmetry of the magnetization with a pulse gradient. So the dipolar Hamiltonian is expressed as (isolated spin Hamiltonian and J coupling Hamiltonian are not included here because they don't contribute to iMQC signals),

$$H_d = \hbar \sum_{i=1}^{N} \sum_{j=1}^{N} D_{ij} (3I_{zi} I_{zj} - \vec{I}_i \vec{I}_j) \tag{0.22}$$

where $Dij = \frac{\mu_0}{4\pi} \frac{\gamma^2 \hbar}{4} \frac{1 - 3\cos^2 \theta_{ij}}{r^3{}_{ij}}$ .

Correction two, or Equation (0.22), which disregard the HTA approximation will also be used here in order to explain the multi-quantum coherence phenomenon. Let me rephrase it here,

$$\sigma_0 = 1 - \frac{\hbar\omega}{kT} \sum_{i=1}^{N} I_{zi} + \frac{1}{2} \left( \frac{\hbar\omega}{kT} \right)^2 \sum_{i,j=1,i \neq j}^{N} I_{zi} I_{zj} \tag{0.23}$$

The following derivation is for water-water coherence. The derivations of water-acetone and acetone-acetone terms are similar. The first 90 degree hard pulse is applied along the x-axis, multi-quantum transitions of different orders are generated from various terms:

$$\begin{aligned} I_{zi} &\xrightarrow{90x} \frac{i}{2} \left( -I_{+i} + I_{-i} \right) \\ I_{zi} I_{zj} &\xrightarrow{90x} \frac{1}{4} \left( -I_{+i} I_{+j} - I_{-i} I_{-j} + I_{-i} I_{+j} + I_{+i} I_{-j} \right) \end{aligned} \tag{0.24}$$

The first term in (0.22) represents single-quantum coherences, the second one represents double quantum and zero quantum coherences. After free-evolution period $\tau$ and the pulse gradient of strength G and length T (the magnetic field gradient orients along with the z-axis and it is constant), the two spin order terms of density matrix becomes:

$$\sigma_1 = \frac{1}{2}\left(\frac{\hbar\omega}{kT}\right)^2 \sum_{i,j=1,i\neq j}^{N} \frac{1}{4}\left[-I_{+i}I_{+j}e^{i(-2\Delta\omega t_1 - 2\gamma GTz)} - I_{-i}I_{-j}e^{i(2\Delta\omega t_1 + 2\gamma GTz)} + I_{-i}I_{+j} + I_{+i}I_{-j}\right] \quad (0.25)$$

where $\Delta\omega$ represents the resonance offsets of I spin, and $\gamma GTz$ is the dephasing angle at the position z attributed to the gradient pulse. The second 90 degree RF converts these operators into (0.26) and no net signal originated from ZQC.

$$\begin{aligned}
I_{-i}I_{+j} &\xrightarrow{(90)x} \frac{i}{2}\left(I_{zi}I_{+j} - I_{+i}I_{zj}\right) \\
I_{+i}I_{-j} &\xrightarrow{(90)x} \frac{i}{2}\left(I_{zi}I_{+j} + I_{+i}I_{zj}\right) \\
I_{-i}I_{+j} &\xrightarrow{(90)x} \frac{i}{2}\left(I_{zi}I_{-j} - I_{-i}I_{zj}\right) \\
I_{+i}I_{-j} &\xrightarrow{(90)x} \frac{i}{2}\left(-I_{zi}I_{-j} + I_{-i}I_{zj}\right)
\end{aligned} \quad (0.26)$$

Then this RF pulse, external magnetic field and the magnetic gradient pulses transforms the density matrix into,

$$\sigma_2 = -\frac{1}{2}\left(\frac{\hbar\omega}{kT}\right)^2 \sum_{i,j=1,i\neq j}^{N} \frac{1}{4}\begin{bmatrix} \frac{i}{2}\left(I_{-i}I_{zj} + I_{zi}I_{-j}\right)e^{i(-2\Delta\omega t_1 - 2\gamma GTz + \Delta\omega t_2 + n\gamma GTz)} \\ -\frac{i}{2}\left(I_{-i}I_{zj} + I_{zi}I_{-j}\right)e^{i(2\Delta\omega t_1 + 2\gamma GTz + \Delta\omega t_2 + n\gamma GTz)} \end{bmatrix} \quad (0.27)$$

In the following, we retained only terms that are observable after evolution under chemical shift Hamiltonian,

$$\sigma_3 = -\frac{1}{2}\left(\frac{\hbar\omega}{kT}\right)^2 \sum_{i,j=1,i\neq j}^{N} \frac{1}{4}\left[\frac{i}{2}\left(I_{-i}I_{zj} + I_{zi}I_{-j}\right)e^{i\left(-2\Delta\omega t_1 - 2\gamma GTz + \Delta\omega t_2 + n\gamma GTz\right)}\right] \tag{0.28}$$

Dipolar coupling transforms the one-quantum operator $I_{-i}I_{zk}$ terms, after one commutation with the dipolar Hamiltonian, into $I_{-i}$ operators, with an observable signal $I_{-i}I_{zj} \xrightarrow{D_{i,j}I_{zi}I_{zj}t_2} -iI_{-i}\sin\left(\frac{D_{i,j}t_2}{2}\right)$. It is clearly shown if n = +2 the signal from double quantum coherence is refocused at t2 = 2t1. From equation (0.28), it is also clear that the zero quantum coherence can be created by the first 90x pulse and would not need further refocusing after the first gradient. The optimization suggests replacing the first 90x pulse to a 45x pulse.

The second explanation is a classical description which uses revised block equation and DDF. In later section 2.7, a distant dipolar field-corrected block equation is shown, and we borrow some conclusions from that. Now we are going to analyze the CRAZED experiment in this classical treatment (based on figure 3).

Before starting this explanation, we would like to introduce some concepts about frequency encoding and modulation. Considering a piece of sample in magnetic field, the sample can be divided to many tiny sections along X-direction. If a linear magnetic field gradient along the X-axis G is introduced into the system, each magnetization M will have a different Larmor frequency $\omega$ changing with distance x. This is called frequency encoding. The accumulated phase of each position, $\phi(\vec{x},T)$, depends on how long the gradient is given,

$$\phi(\vec{x},T) = -\gamma\vec{x}\int_0^T G(\vec{x},t)dt \tag{0.29}$$

Thus, each magnetization has its own phase. If the magnetization vector of each position could be detected and plotted, a helix along the X-axis would be seen. This magnetization is now called "modulated" and its modulation can be described as a spatial frequency in k-space.

$$\vec{k}(T) = \frac{\gamma}{2\pi}\int_0^T G(\vec{x},t)dt \tag{0.30}$$

In the iDQC sequence, the first 90 degree puts the equilibrium magnetizations of water and acetone on the transverse plane. During the $t_1$ period, the effective distant dipolar field is zero because the longitudinal magnetization is zero. At the end of the first X-gradient, the magnetization of water is:

$$\begin{aligned} &M_{Z,W} = 0 \\ &M_W^+ = M_{X,W} + iM_{Y,W} = -iM_{0,W}\exp\left[i\left(\Omega_W\left(t_1+T\right)+\gamma GTx\right)\right] \end{aligned} \tag{0.31}$$

The transverse magnetization carries an X-modulation. This modulation is transferred to longitudinal magnetization after the second 90x pulse:

$$\begin{aligned} M_{Z,W} &= -M_{0,W}\cos\left[\Omega_W\left(t_1+T\right)+\gamma GTx\right] \\ M_W^+ &= M_{0,W}\sin\left[\Omega_W\left(t_1+T\right)+\gamma GTx\right] \end{aligned} \tag{0.32}$$

The transverse magnetization water then evolves under the bulk magnetic field, the second X-gradient and DDF. At the end of the d2 delay, the water magnetization is,

$$\begin{aligned} M_W^+ &= \exp\left[i\Omega_W\left(d_2+2T\right)\right]\cdot\exp\left[i\gamma G2Tx\right] \\ &\times\exp\left[i\gamma\left(B_{DDF,W}^{eff}+B_{DDF,A}^{eff}\right)\left(d_2+2T\right)\right] \\ &\times M_{0,W}\sin\left[\Omega_W\left(t_1+T\right)+\gamma GTx\right] \end{aligned} \tag{0.33}$$

By using the homo-molecular and hetero-molecular equations from section 2.7, the water transverse magnetization can then be further separated to homo-molecular and hetero-molecular DDF evolutions (redefine d2 + 2T -> d2 and t1 + T -> t1 because of d2 >> 2T and t1 >> T),

$$\begin{aligned} M_W^+ &= \mathrm{e}^{i\Omega_W d_2}\cdot\mathrm{e}^{i\gamma G2Tx}\cdot\exp\left[i\gamma\left(\Delta_s\mu_0 M_{Z,W}+\Delta_s\mu_0\frac{2}{3}M_{Z,A}\right)d_2\right] \\ &\times M_{0,W}\sin\left[\Omega_W\left(t_1+T\right)+\gamma GTx\right] \end{aligned} \tag{0.34}$$

Substituting the $M_{Z,W}$ and $M_{Z,A}$ with $-M_{0,W}\cos\left(\Omega_W t_1+\gamma GTx\right)$ and $-M_{0,A}\cos\left(\Omega_A t_1+\gamma GTx\right)$, respectively, the relationship between the distant dipolar field and the evolution during t1 is emerging,

$$\begin{aligned} M_W^+ &= \mathrm{e}^{i\Omega_W d_2}\cdot\mathrm{e}^{i\gamma G2Tx}\cdot M_{0,W}\sin\left[\Omega_W\left(t_1+T\right)+\gamma GTx\right] \\ &\times\exp\left[-i\gamma\Delta_s\mu_0\left(M_{0,W}\cos\left(\Omega_W t_1+\gamma GTx\right)+\frac{2}{3}M_{0,A}\cos\left(\Omega_A t_1+\gamma GTx\right)\right)d_2\right] \end{aligned} \tag{0.35}$$

We now introduce the Bessel function, $\mathrm{e}^{iA\cos B} = \sum_{m=-\infty}^{\infty} i^m \cdot J_m(A) \cdot e^{imB}$. Using Bessel function expansions, equation (0.30) can be written as,

$$
\begin{aligned}
M_W^+ &= \mathrm{e}^{i\Omega_W d_2} \cdot \mathrm{e}^{i\gamma G2Tx} \cdot M_{0,W} \sin\left[\Omega_W t_1 + \gamma GTx\right] \\
&\times \sum_{m=-\infty}^{\infty} i^m \cdot J_m\left(-\gamma \Delta_s \mu_0 d_2 M_{0,W}\right) \cdot e^{im(\Omega_W t_1 + \gamma GTx)} \\
&\times \sum_{m'=-\infty}^{\infty} i^{m'} \cdot J_{m'}\left(-\gamma \Delta_s \mu_0 d_2 \frac{2}{3} M_{0,A}\right) \cdot e^{im'(\Omega_A t_1 + \gamma GTx)} \\
&= \mathrm{e}^{i\Omega_W d_2} \cdot \frac{M_{0,W}}{2i} \cdot \left[e^{i(\Omega_W t_1 + \gamma GTx + \gamma G2Tx)} - e^{-i(\Omega_W t_1 + \gamma GTx - +\gamma G2Tx)}\right] \\
&\times \sum_{m=-\infty}^{\infty} i^m \cdot J_m\left(-\gamma \Delta_s \mu_0 d_2 M_{0,W}\right) \cdot e^{im(\Omega_W t_1 + \gamma GTx)} \\
&\times \sum_{m'=-\infty}^{\infty} i^{m'} \cdot J_{m'}\left(-\gamma \Delta_s \mu_0 d_2 \frac{2}{3} M_{0,A}\right) \cdot e^{im'(\Omega_A t_1 + \gamma GTx)} \\
&= \mathrm{e}^{i\Omega_W d_2} \cdot \frac{M_{0,W}}{2i} \cdot \left[e^{i(\Omega_W t_1 + \gamma 3GTx)} - e^{-i(\Omega_W t_1 - \gamma GTx)}\right] \\
&\times \sum_{m=-\infty}^{\infty} i^m \cdot J_m\left(-\frac{\Delta_s d_2}{\tau_{d,W}}\right) \cdot e^{im(\Omega_W t_1 + \gamma GTx)} \\
&\times \sum_{m'=-\infty}^{\infty} i^{m'} \cdot J_{m'}\left(-\frac{2\Delta_s d_2}{3\tau_{d,A}}\right) \cdot e^{im'(\Omega_A t_1 + \gamma GTx)}
\end{aligned}
\qquad (0.36)
$$

where $\tau_{d,W} = \left(\gamma \mu_0 M_{0,W}\right)^{-1}$ and $\tau_{d,W} = \left(\gamma \mu_0 M_{0,A}\right)^{-1}$ are the "dipolar demagnetization times" of water and acetone, respectively. Since there are gradient terms, the average transverse magnetization in equation (0.36) would vanish unless the combination of m and m' eliminates the gradient terms. Most such combinations are related to high order terms and can be ignored. Only four considerable combinations survive: m = -3 or m = -1 with m′ = 0; and m = -2 or m = 0 with m′ = -1. Equation (0.36) then becomes,

$$
\begin{aligned}
M_W^+ &= e^{i\Omega_W d_2}\cdot\frac{M_{0,W}}{2i}\cdot J_0\left(-\frac{2\Delta_s d_2}{3\tau_{d,A}}\right)\cdot\left[i^{-3}J_{-3}\left(-\frac{\Delta_s d_2}{\tau_{d,W}}\right)e^{-i2\Omega_W t_1}-i^{-1}J_{-1}\left(-\frac{\Delta_s d_2}{\tau_{d,W}}\right)e^{-i2\Omega_W t_1}\right] \\
&+e^{i\Omega_W d_2}\cdot\frac{M_{0,W}}{2i}\cdot i^{-1}\cdot J_{-1}\left(-\frac{2\Delta_s d_2}{3\tau_{d,A}}\right)\cdot\left[i^{-2}J_{-2}\left(-\frac{\Delta_s d_2}{\tau_{d,W}}\right)e^{-i(\Omega_W+\Omega_A)t_1}-i^{0}J_{0}\left(-\frac{\Delta_s d_2}{\tau_{d,W}}\right)e^{-i(\Omega_W+\Omega_A)t_1}\right] \\
&= e^{-i2\Omega_W t_1}\, e^{i\Omega_W d_2}\cdot\frac{M_{0,W}}{2i}\cdot i^{-3}\cdot J_0\left(-\frac{2\Delta_s d_2}{3\tau_{d,A}}\right)\cdot\left[J_{-3}\left(-\frac{\Delta_s d_2}{\tau_{d,W}}\right)-i^{-1}J_{-1}\left(-\frac{\Delta_s d_2}{\tau_{d,W}}\right)\right] \\
&+e^{-i(\Omega_W+\Omega_A)t_1}\, e^{i\Omega_W d_2}\cdot\frac{M_{0,W}}{2i}\cdot i^{-3}\cdot J_{-1}\left(-\frac{2\Delta_s d_2}{3\tau_{d,A}}\right)\cdot\left[J_{-2}\left(-\frac{\Delta_s d_2}{\tau_{d,W}}\right)-J_{0}\left(-\frac{\Delta_s d_2}{\tau_{d,W}}\right)e^{-i(\Omega_W+\Omega_A)t_1}\right]
\end{aligned}
\qquad (0.37)
$$

A Bessel function is an even function when its order is even and an odd function when its order is odd. It also follows the rules below,

$$
\begin{aligned}
J_{-N}(B) &= (-1)^N J_N(B) \\
J_N(B) &= \frac{B}{2N}\left[J_{N-1}(B)+J_{N+1}(B)\right]
\end{aligned}
\qquad (0.38)
$$

The un-vanished water magnetization now is,

$$
M_W^+ = M_{0,W}e^{i\Omega_W d_2}
$$

$$
\times\left[e^{-i2\Omega_W t_1}2\left(\frac{\tau_{d,W}}{\Delta_S d_2}\right)J_2\left(\frac{\Delta_S d_2}{\tau_{d,W}}\right)J_0\left(\frac{2\Delta_S d_2}{3\tau_{d,A}}\right)+e^{-i(\Omega_W+\Omega_A)t_1}\left(\frac{\tau_{d,W}}{\Delta_S d_2}\right)J_1\left(\frac{\Delta_S d_2}{\tau_{d,W}}\right)J_1\left(\frac{2\Delta_S d_2}{3\tau_{d,A}}\right)\right]
$$

(0.39)

Following the same derivation, the un-vanished acetone magnetization is,

$$M_A^+ = M_{0,A} e^{i\Omega_A d_2}$$
$$\times\left[e^{-i2\Omega_A t_1} 2\left(\frac{\tau_{d,A}}{\Delta_S d_2}\right) J_2\left(\frac{\Delta_S d_2}{\tau_{d,A}}\right) J_0\left(\frac{2\Delta_S d_2}{3\tau_{d,W}}\right) + e^{-i(\Omega_W+\Omega_A)t_1}\left(\frac{\tau_{d,A}}{\Delta_S d_2}\right) J_1\left(\frac{\Delta_S d_2}{\tau_{d,A}}\right) J_1\left(\frac{2\Delta_S d_2}{3\tau_{d,W}}\right)\right]$$

(0.40)

Where the $\Delta_S$ here is -1 (X-modulation). Water and acetone would evolve at their own frequencies, $\Omega_W$ and $\Omega_A$, during the following acquisition period t2. Each magnetization carries two phases from t1 evolution. Thus, four peaks would be on the final 2D spectrum, $2W(-\Omega_w - \Omega_w, \Omega_w)$, $AW(-\Omega_A - \Omega_w, \Omega_w)$, $WA(-\Omega_A - \Omega_w, \Omega_A)$ and $2A(-\Omega_A - \Omega_A, \Omega_A)$. Equation (0.34) and (0.35) agree with the X-crazed spectrum (figure 3) not only in the peak positions but also in the peak intensities. In the limit of B << 1, Bessel functions $J_0(B) \approx 1$, $J_1(B) \approx \frac{B}{2}$ and $J_2(B) \approx \frac{B^2}{8}$, the peaks 2W, AW, WA and 2A would be approximately proportional to $\Delta_S d_2 (M_{0,W})^2$, $\Delta_S d_2 \frac{2}{3}(M_{0,A})(M_{0,W})$, $\Delta_S d_2 \frac{2}{3}(M_{0,W})(M_{0,A})$ and $\Delta_S d_2 (M_{0,A})^2$, respectively.

## 2.6 Description for CRAZED sequences

Double quantum and zero quantum coherences can be visualized by drawing the energy level diagram for a 2-spin system (figure 4.b). Double quantum coherences correspond to simultaneous transitions of both spins in the same direction (a flip-flip transition or up-up to down-down). The net change in angular momentum is 2 (instead of 1 for a standard transition). The zero quantum coherence is the simultaneous transition of two spins in opposite directions (a flip-flop transition, or up-down to down-up). The net change in angular momentum is 0.

The energy (or frequency) for any transition is given by the difference in the energy levels; thus, the frequency for a double quantum transition is the energy difference between the upper energy level and the lower energy level (E4-E1). The energy of the uppermost level is given by $E4 = -\omega_1/2 - \omega_2/2$, and the lower energy level is $E1 = \omega_1/2 + \omega_2/2$. The energy of the transition therefore comes at the sum of the two frequencies. Since a zero quantum transition is a transition between energy levels E2 and E3, the energy of that transition is $E2 - E3 = \omega_1 - \omega_2$.

The effect of the gradients is slightly more complex because it acts on the system in two ways. First, it works as a coherence selection gradient. When the gradient is applied to the system, it creates a distribution of resonance frequencies depending on spatial position. This changes the effective magnetic field at different locations and causes the energy levels to shift depending on the spatial position of the spins. Since a double quantum transition occurs when both spins flip in the same direction, the effect of the gradient is doubled because it contributes to the resonance frequency once for each spin. For the zero quantum coherence, the spins flip in opposite directions and the effect of the gradient is canceled.

In addition, the gradient breaks the magnetic isotropy of the sample. When combined with the mixing pulse, the gradient introduces the dipolar field to the sample and converts the unobservable multiple quantum terms into observable single quantum terms. A more visual explanation of how this works is given in figure 4.c. After the application of the 90° pulse, all the magnetization is pointed along one direction in the transverse plane. The gradient is applied, which winds the magnetization into a helix along the direction of the gradient. The second pulse tips some of that magnetization back along the z-axis. Depending on the phase of the magnetization vector before the second pulse, the z-component of the magnetization will be pointed either along +z or –z. The z-magnetization created by this second pulse exerts a force on the remaining magnetization in the transverse plane, causing it to refocus. The time required for the magnetization to refocus depends on the size of the magnetization that was tipped along the z axis, and this time is referred to as the "dipolar demagnetizing time". The concept of refocusing created by the gradient and the pulse provides a qualitative description of the behavior of the dipolar field and how it transforms the multiple quantum signal to observable signal.

## 2.7 The distant dipolar field – corrected Bloch equations

In previous sections, we reviewed how the traditional dipolar coupling works and how CRAZED experiments challenge it. After understanding it from NMR experiments, we are going back to the theory and try to conclude a corrected theory.

As described above, the spatial integral of the distant dipole-dipole couplings can be processed as an additional magnetic field $\left(\vec{B}_{DDF}\right)$, which is created by the sample itself. Using the water/acetone system as an example, the DDF-corrected Bloch equation of the water can be written as (ignore flow effects in the case of constant velocity $\left\langle \vec{\upsilon}\left(\vec{r}\right),\nabla\vec{M}\right\rangle$ and magnetic field created by radiation damping $\vec{B}_{RD}\left(\vec{r}\right)$):

$$\begin{aligned}\frac{\partial\vec{M}}{\partial t}&=\gamma\vec{M}\times\vec{B}+D\nabla^2\vec{M}-\frac{M_x\hat{x}+M_y\hat{y}}{T_2\left(\vec{r}\right)}+\frac{M_0-M_z}{T_1\left(\vec{r}\right)}\hat{z}\\ \vec{B}&=\frac{\Omega\left(\vec{r}\right)}{\gamma}\hat{z}+G\left(\vec{r}\cdot\vec{s}\right)\hat{z}+\vec{B}_{DDF}\left(\vec{r}\right)\end{aligned} \tag{0.41}$$

where D is the diffusion constant. In most cases it is not possible to obtain analytical solutions to the Bloch equations. Problems arise when specific boundary conditions need to be enforced, or when arbitrary distributions of relaxation times and resonance frequency offsets need to be modeled.

Since the sample is modulated, the sample wide magnetization is replaced by the magnetization of each position. Because diffusion and relaxation are not the reasons for the CRAZED spectrum above, we will ignore them here. Even though the water in this system is concentrated enough to cause radiation damping in itself, the two asymmetric gradients before and after the second 90-degree pulse make the sample completely modulated so that its average transverse magnetization is zero right after the second gradient. Thus, we can also ignore the radiation damping here and focus only on the terms related to the bulk magnetic field $\Omega(\vec{r})$, the gradients $G(\vec{r}\cdot\vec{s})$ and the DDF $\vec{B}_{DDF}(\vec{r})$. The form of $\vec{B}_{DDF}(\vec{r})$ should be equation 0.41 which is only the secular part.

The DDF (equation 0.41) can be further sorted into contributions of two types: homo-molecular and hetero-molecular. In the view of water, for example, the homo-molecular DDF is caused by the DD couplings between water molecules, while the hetero-molecular DDF is from DD couplings between water and other molecules, which is only acetone in this system. The two distant dipolar fields seen from water are:

$$\begin{aligned} \vec{B}_{DDF,w}(\vec{r}) &= \iiint d^3\vec{r}\,'\times\frac{1-3\cos^2\theta_{rr'}}{2\left|\vec{r}-\vec{r}\,'\right|^3}\times\left[3\vec{M}_{z,w}(\vec{r}\,')\hat{z}-\vec{M}_w(\vec{r}\,')\right] \qquad Homo- \\ \vec{B}_{DDF,a}(\vec{r}) &= \iiint d^3\vec{r}\,'\times\frac{1-3\cos^2\theta_{rr'}}{2\left|\vec{r}-\vec{r}\,'\right|^3}\times 2\vec{M}_{z,a}(\vec{r}\,')\hat{z} \qquad Hetero- \end{aligned} \tag{0.42}$$

The hetero-molecular DDF doesn't contain any transverse component because $\left|\Omega_w-\Omega_a\right| = \gamma B_{DDF,a}$.

The spatial integral makes it unrealistic to calculate the real DDF effect even with today's computational power. Fortunately, Deville et al. pointed out in 1979 that the calculation of the DDF can be greatly simplified for the highly modulated situation [28]. This observation considerably speeds up numerical calculations of the dipolar field. This elegant method takes advantage of the fact that, while equations (0.42) are non-local in real space, they are local in k-space, the spatial frequency domain. In k-space, the magnetization and distant dipolar field are:

$$\begin{aligned}\vec{M}\left(\vec{k}\right)&=\iiint d^3\vec{r}\times e^{i\vec{k}\cdot\vec{r}}\times\vec{M}\left(\vec{r}\right)\\ \vec{B}_{DDF}\left(\vec{k}\right)&=\iiint d^3\vec{r}\times e^{i\vec{k}\cdot\vec{r}}\times\vec{B}_{DDF}\left(\vec{r}\right)\end{aligned}\tag{0.43}$$

Using Fubini's theorem and complicated derives, we can reach the conclusion:

$$\vec{B}\left(\vec{k}\right)=\left[3M_z\left(\vec{k}\right)\hat{z}-\vec{M}\left(\vec{k}\right)\right]\left[1-3\left(\hat{k}\cdot\hat{z}\right)^2\right]\tag{0.44}$$

If the modulation $\vec{k}$ is only along one direction, the distant dipolar field in k-space is just a constant multiplying the magnetization in k-space. It implies that, with a modulation introduced along only one direction, the effective distant dipolar field is local in real space:

$$\vec{B}_{DDF}\left(\vec{r}\right)=\left[3M_z\left(s\right)\hat{z}-\vec{M}\left(s\right)\right]\cdot\left[3\left(\hat{s}\cdot\hat{z}\right)^2-1\right]\tag{0.45}$$

where s is the component of $\vec{r}$ along $\hat{s}$. Thus, the effective distant dipolar field, both homo-molecular and hetero-molecular, seen from water can be further simplified from equation (0.42) to:

$$
\begin{aligned}
\vec{B}^{eff}_{DDF,w}(\vec{r}) &= \left[3(\hat{s}\cdot\hat{z})^2 - 1\right]\left[3M_{z,w}(\vec{r})\hat{z} - \vec{M}_w(\vec{r})\right] \\
&= \left[3(\hat{s}\cdot\hat{z})^2 - 1\right]3M_{z,w}(\vec{r})\hat{z} \quad Homo- \\
\vec{B}^{eff}_{DDF,a}(\vec{r}) &= \left[3(\hat{s}\cdot\hat{z})^2 - 1\right]2M_{z,a}(\vec{r})\hat{z} \qquad Hetero-
\end{aligned} \tag{0.46}
$$

where the $\vec{M}_w(\vec{r})$ has been removed because $\vec{M}_w(\vec{r})\times\vec{M}_w(\vec{r}) = 0$, removing the $\vec{M}_w(\vec{r})$ contribution from the Bloch equation. Equations (0.46) tell us that, the effective DDF generated by a Z-axis modulation is twice as strong and of the opposite sign to the effective DDF by X or Y modulation. This leads to a very important application, anisotropy mapping which will be discussed later.

# 3. Introduction to Diffusion MRI and Susceptibility MRI

Diffusion MRI is a series of medical imaging methods which allows of the mapping of diffusion process of molecules, mainly water in biological system. In this category, diffusion tensor imaging (DTI) is a very unique technique for studying water motion in a tensor way [29]. Another medical imaging method focused on tensor is susceptibility tensor imaging (STI), it is a novel technique to measure and quantify the extensive anisotropic magnetic susceptibility in biologic tissues, such as white matter of the central nervous system [30-33].

## 3.1 Diffusion MRI

D Le Bihan and E Breton, et al [34] discovered diffusion MRI techniques and shown corresponding images on normal and diseased brain in 1985. After that, diffusion MRI also referred to as diffusion tensor imaging (DTI) has been extremely successful in clinical and research. It can be applied to study neurological disorders [35], detect acute stroke and obtain perfusion imaging [36], et al.

### 3.1.1 Diffusion

We define the motion of water molecules as three categories, bulk motion, flow and diffusion [29]. The bulk motion means the water molecules movement is more than a pixel of NMR experiments in dimension. The flow is defined as one-directional water motion within a pixel of NMR experiments. This flow can affect MR signal, especially phase significantly. In non-unidirectional flow and large blood flow due to convoluted capillary structures, the flow is always discarded because the population of water within blood vessels of the brain is small (about 5%) compared to those in the parenchyma. What we are interested in is diffusion (random motion or Brownian motion). Any water molecule in a certain pixel will spread out according to a "Gaussian distribution".

The NMR pixel size is typically 2 – 5 mm, while the actual amount of water diffusion is approximately 5 – 10 $\mu m$ during NMR measurements. We need to sensitize the signal intensity to this amount of water diffusion or diffusion constant. This is how we measure diffusion.

Now let's look at the mathematical model of diffusion. Given the concentration $\rho$ and flux $J$, a relationship between the flux and the concentration gradient is given by Fick's first law:

$$J(x,t) = -D \cdot \nabla \rho(x,t) \tag{0.47}$$

where D is the diffusion coefficient. Then, given conservation of mass, the continuity equation relates the time derivative of the concentration with the divergence of the flux:

$$\frac{\partial \rho(x,t)}{\partial t} = -\nabla \cdot J(x,t) \tag{0.48}$$

Putting the two together, we get the diffusion equation:

$$\frac{\partial \rho(x,t)}{\partial t} = D \cdot \nabla^2 \rho(x,t) \tag{0.49}$$

### 3.1.2 Diffusion weighted imaging (DWI)

Let's revisit the equation (0.41). If we don't consider the bulk movement and the flow of water molecules, equation (0.41) is simplified to:

$$\frac{d\vec{M}}{dt} = \gamma \vec{M} \times \vec{B} + \nabla \cdot D \nabla \vec{M} - \frac{M_x \hat{x} + M_y \hat{y}}{T_2(\vec{r})} + \frac{M_0 - M_z}{T_1(\vec{r})} \hat{z} \tag{0.50}$$

where D is the diffusion tensor. H.C. Torrey mathematically showed how the Bloch equations for magnetization would change with the addition of diffusion [7]. He modified Bloch's original description of transverse magnetization to include diffusion terms and the application of a spatially varying gradient.

Consider the simplest case which is for isotropic diffusion, the diffusion tensor D can be rewritten as a multiplication of a scalar D and an identity matrix. Then the Block-Torrey equation will have the solution:

$$M = M_{bloch} e^{-\frac{1}{3}\gamma^2 G^2 t^3} = e^{-bD_0} \tag{0.51}$$

For anisotropic diffusions, the attenuation should be:

$$A = e^{-\sum_{i,j} b_{ij} D_{ij}} \tag{0.52}$$

where the $b_{ij}$ terms incorporate the gradient fields in three directions. By applying different gradient fields in different directions, we can solve for diffusion coefficients by fitting the data.

Theoretical model has been introduced, and then we need to think about how to achieve all of these by MRI sequences.

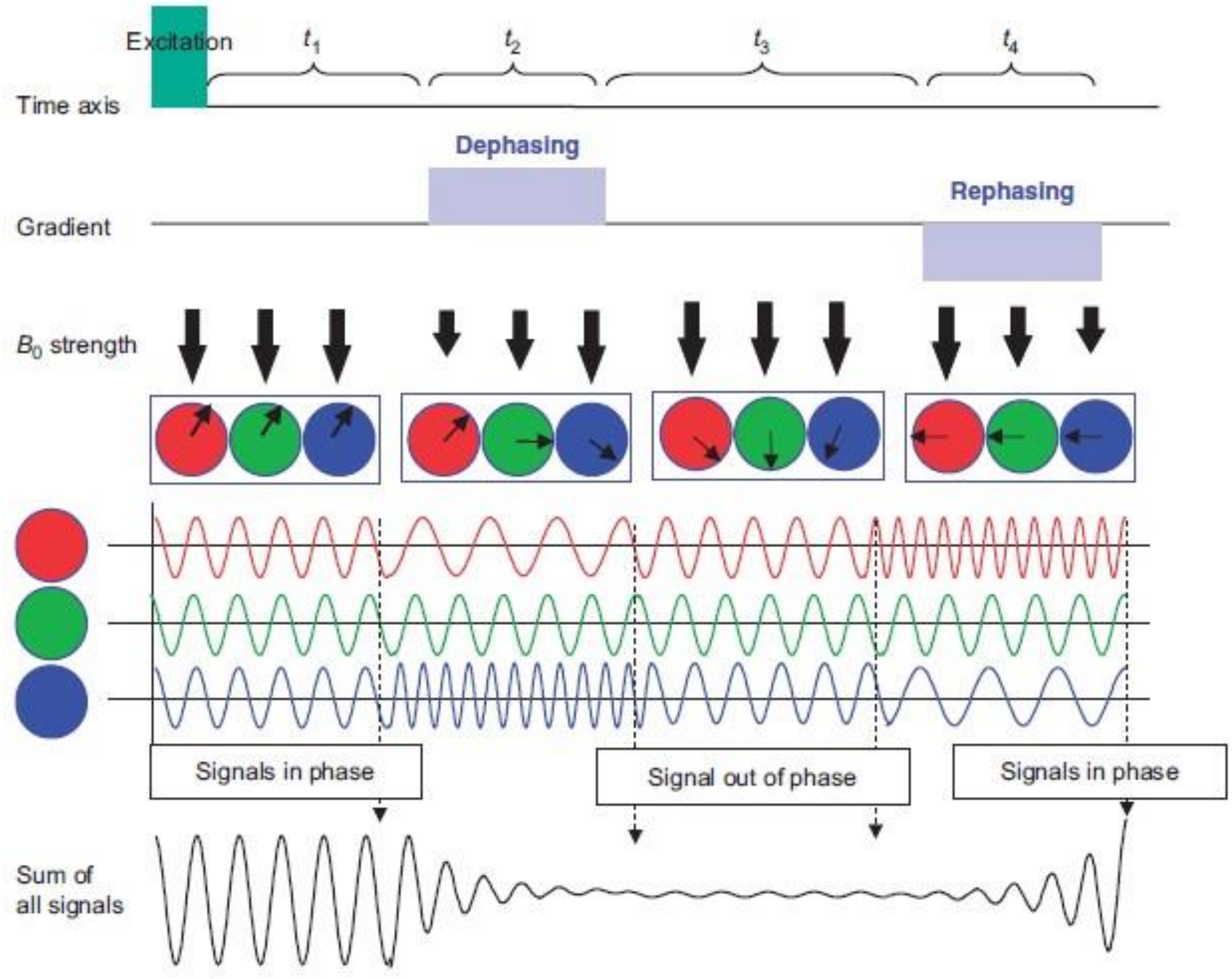

**Figure 5: An example of a dephase-rephrase experiment by gradient application. Red, green, and blue circles indicate three water molecules located at different positions in a sample tube. Thick arrows indicate the strengths of magnetic**

**field strength ($B_0$), and narrow arrows indicate phases of MR signals from each molecule [29].**

As what is shown in figure 5, a pair of positive and negative gradients is applied. After excitation RF pulse (time t1), protons at different locations start to give MR signals at the same frequency. During the first gradient application (t2), protons start to see different $B_0$ and resonate at different frequencies, depending on their locations. In t3 period, the phases of the signals are no longer identical among the protons. A overall signal loss happens, so the first gradient is called the “dephasing” gradient. The second gradient (t4), called the rephasing gradient, has the opposite polarity. If the strength and length of the rephasing gradient are identical to those of dephasing one, nothing has happened in the past except that the resultant signal is sensitized to diffusion. This is because perfect refocusing happens only when water molecules do not change their locations in between the applications of the two dephase-rephase gradients. If water moves, it results in disruption of the phase gradient across the sample. MR cannot measure the phase of individual water molecules, but it can detect the imperfect rephrasing by the loss of signal intensity [29]. Three important points needed to clarify, the diffusion measurement is noninvasive and does not require injection of any chemical tracers; it measures the water motion along a predetermined axis referring to the direction of field gradient applied; flow or bulk motions actually lead to different outcomes in this experiment because they result in perfect refocusing.

Another parameter in diffusion measurement, it is the length of the molecular displacement during the measurement is typically 1 – 20 um, depending on sample, temperature, and pulse sequence.

### 3.1.3 Diffusion tensor imaging (DTI)

As the last three points mentioned in the last section, we actually can measure diffusion along any axis if the field gradient is putting on that axis. Usually we don't expect free water diffusion has this directionality, while it has when we measure it inside a living system.

Diffusion directionality and flow are different. Suppose we drop ink into a media, the center of the ink moves when there is a flow. While the media freely diffuses and there is no flow, the shape of ink becomes a sphere and the center of it doesn't move. We call it "isotropic" diffusion. If the shape of the ink becomes oval or ellipsoid while the center of it still stays, this type of diffusion is called "anisotropic" diffusion. This kind of diffusion always happens in biological tissues because the water tends to diffusion in a preferential axis in it. Apparently, we cannot describe this type of diffusion process using a single diffusion measurement of by a single diffusion constant.

The anisotropic diffusion can give us much information about the underlying anatomical architecture of living tissues, such as axonal tracts in nervous tissues or protein filaments in muscle. Using diffusion tensor imaging (DTI), we can access this precious information.

Fig.6 clearly shows that six parameters are needed to define an ellipsoid. $\alpha$, $\beta$ and $\gamma$ decide the diffusion orientation, namely $\vec{v}_1$, $\vec{v}_2$ and $\vec{v}_3$. $\lambda_1$, $\lambda_2$ and $\lambda_3$ decide the diffusion strength in each orientation.

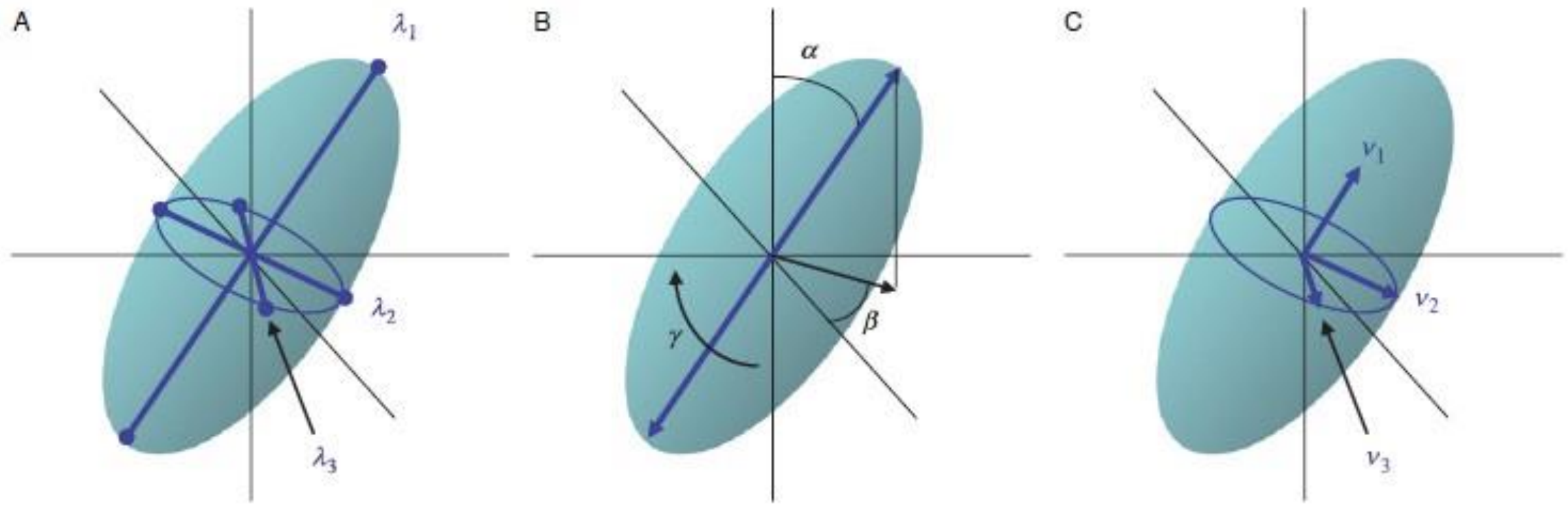

**Figure 6: Parameters needed to define a 3D ellipsoid.**

Naturally, our task is to determine the six parameters $(\lambda_1 \quad \lambda_2 \quad \lambda_3 \quad \vec{v}_1 \quad \vec{v}_2 \quad \vec{v}_3)$. What we do is to measure diffusion constants along multiple independent axes, in order to accurately define the ellipsoid under the existence of measurement errors.

$$\overline{\overline{D}} = \begin{bmatrix} D_{xx} & D_{xy} & D_{xz} \\ D_{yx} & D_{yy} & D_{yz} \\ D_{zx} & D_{zy} & D_{zz} \end{bmatrix} \xrightarrow{diagonalization} \lambda_1 \quad \lambda_2 \quad \lambda_3 \quad v_1 \quad v_2 \quad v_3 \qquad (0.53)$$

The diffusion tensor, $\overline{\overline{D}}$, is a symmetric tensor, which means $D_{ij} = D_{ji}$, and, thus, there are six independent parameters, which makes sense because it intrinsically contains the six parameters of the diffusion ellipsoid.

Once we obtain the six parameters of the diffusion ellipsoid at each pixel, our next task is to visualize it so that we can appreciate the neuroanatomy. The most complete way of doing this is to place the 3D ellipsoid at each pixel. However, this is not really a practical method for routine use. Unfortunately, our eyes (or the computer screen) can effectively appreciate (or display) only images with 8-bit (256) grayscale (i.e., pixel intensity) or 24-bit (red/green/blue, RGB) color presentation. Therefore, it becomes important to reduce the six-parameter/pixel information to 8-bit grayscale or 24-bit color presentation. Another problem is about how to present the anisotropy. One of the most widely used indices, fractional anisotropy (FA), is:

$$FA = \sqrt{\frac{1}{2}} \frac{\sqrt{(\lambda_1 - \lambda_2)^2 + (\lambda_2 - \lambda_3)^2 + (\lambda_3 - \lambda_1)^2}}{\sqrt{\lambda_1^2 + \lambda_2^2 + \lambda_3^2}} \quad (0.54)$$

So far, scalar values derived from eigenvalues, FA, is easy to produce and have some useful information. On the other hand, orientation information is less straightforward to visualize, quantify, and interpret. In many studies, we discard $\vec{v}_2$ and $\vec{v}_3$ and concentrate on the orientation of the vector $\vec{v}_1$, which is assumed to represent the local fiber orientation.

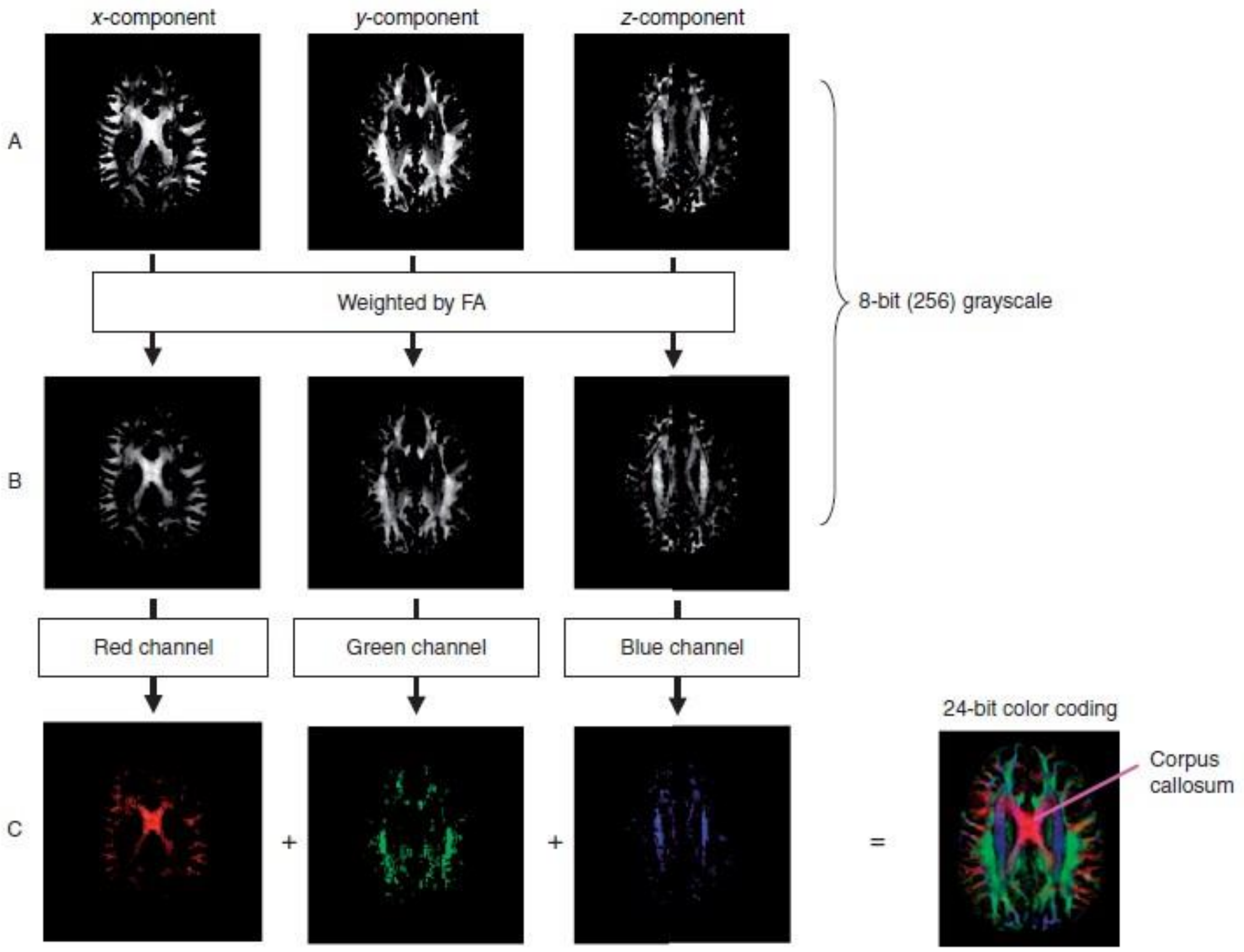

**Figure 7: Steps to create color-coded fiber ($\vec{v}_1$) orientation maps. The x, y, and z components are obtained from the unit vector $\vec{v}_1$ ($\vec{v}_1 = [x, y, z]$).**

One of the most popular ways to visualize the orientation information is with color-coded maps (fig 7). The $\vec{v}_1$ is a unit vector. It consists of x, y, and z components ($\vec{v}_1 = [x, y, z]$) that fulfills $x^2 + y^2 + z^2 = 1$ and each scaled within the 0-1 range. These x, y, and z components can be presented separately using the grayscale as shown above, in which 256 (8 bit) steps of a grayscale is assigned to each vector component. To make low-anisotropy regions where there are supposed to be no dominant fibers, there vector component images can be multiplied by an anisotropy map such FA, which produces cleaner and more informative images. To better visualize fiber orientations in one image, a 24-bit color presentation, which used RGB (8 bit each for red, green, and blue) channels, has been postulated. Namely, the x, y, and z component images are assigned to three RGB principal colors and combined to make one color-coded map.

## 3.2 Susceptibility MRI

There are many susceptibility MRI techniques developed recently, such as quantitative susceptibility mapping (QSM), susceptibility weighted imaging (SWI) and susceptibility tensor imaging (STI), etc.

Magnetic susceptibility is defined as the magnetic response of a substance when it is placed in an external magnetic field [32]. We describe the induced magnetization M as $M = \chi H$. After electromagnetics derivations and the requirement $\chi \ll 1$ for linear materials, we obtain $M = \chi B/\mu_0$ [32]. This formula shows that the induced magnetization is directly proportional to the induced magnetic field $B = B0 + \Delta B$.

Generally, susceptibility related imaging techniques use a 3D multi-echo gradient recalled, fully flow compensated, RF spoiled and high resolution pulse sequence to acquire images [37]. We are dealing with the phase images. However, phase unwrapping should be applied before doing anything. This first step is very necessary for raw phase images produced by each echo, especially for late echoes because they are wrapped very heavily. Laplacian Phase Unwrapping is a useful and specific method to unwrap MR phase images of brains [38-40].

By theoretic derivation, the phase [32] can be denoted as:

$$\begin{aligned}\Delta\varphi &= -\gamma\cdot\Delta\mathrm{B}\cdot t = -\gamma\cdot\Delta\chi\cdot\mathrm{B}_0\cdot t\\ \Delta\mathrm{B} &= \Delta\mathrm{B}_{\text{local geometry}} + \Delta\mathrm{B}_{\text{chemcial shift}} + \Delta\mathrm{B}_{\text{global geometry}} + \Delta\mathrm{B}_{\text{main field}}\end{aligned} \tag{0.55}$$

$\gamma$ is the gyromagnetic constant which is equal to $2\pi\cdot 42.58\,\mathrm{MHz/T}$ for protons. $\Delta\chi$ is the difference of the local magnetic susceptibility of the tissue of interest from its surroundings. $\Delta\mathrm{B}$ represents the local field deviation caused by iron, for example, and $\mathrm{B}_0$ is the field strength.

To show the susceptibility information better, the background noise which includes effects from receiver coil, objects outside the FOV, and objects inside the FOV, and etc. should be removed essentially. Firstly, the phase images can be high pass (HP) filtered to remove some unwanted artifacts [31, 39]. It removes the last 2 terms $\Delta\mathrm{B}_{\text{global geometry}}$ and $\Delta\mathrm{B}_{\text{main field}}$ in equation 1.39. They tend to have a low spatial-frequency dependence (the phase varies slowly over the image) [32, 41]. Secondly, polynomial fitting to the data produces macroscopic phase which should be subtracted from the original phase data [42]. Then the background harmonic phase can be removed using sphere mean value filtering followed by a deconvolution operation to restore the low frequency local phase [38, 43, 44]. At last, projection onto dipole field also can be used to decompose the background field [45].

After all the background phase removal steps, a mask is created from the final phase image by mapping all values above 0 radians to be 1 and linearly or functionally mapping values from –pi to 0 radians to range from 0 to1, respectively [31, 46]. In the end, the susceptibility weighted imaging is the product of magnitude data and the N-th power of phase data, which N is an optimized number.

In general, magnetic susceptibility is direction-dependent. So mathematically saying, magnetic susceptibility can be described by a 3X3 matrix whose elements are denoted as $\chi_{ij}$ [47]. The tensor should be diagonal with equal diagonal elements considering isotropic susceptibility. At the observed nucleus which is assumed to be within a small sphere of Lorentz, $\vec{\mathrm{B}} = (\vec{\mathrm{I}} - \sigma\vec{\mathrm{I}} + \vec{\chi}/3)(\vec{\mathrm{H}} + \vec{\mathrm{h}})$ [47-49]. Here, σ is the chemical shift caused by the screening effect in the electronic shell, $\vec{\mathrm{h}}$ is the demagnetizing field of the object. Using Maxwell's equations to solve for $\vec{\mathrm{h}}$, we found that:

$$\vec{\mathrm{h}} = -\mathrm{FT}^{-1}\left\{\vec{\mathrm{k}}\frac{\vec{\mathrm{k}}^{\mathrm{T}} \cdot \mathrm{FT}\{\chi\} \cdot \vec{\mathrm{H}}}{k^2}\right\} \tag{0.56}$$

By plugging back $\vec{\mathrm{h}}$, we solve for the off-resonance field $\Delta\vec{\mathrm{B}}$, referenced to $(1\text{-}\sigma)\vec{\mathrm{H}}$. Because that the change in phase between 2 tissues after a period of time t can be written as $\Delta\varphi = -\gamma \cdot \Delta\vec{\mathrm{B}} \cdot \mathrm{t}$ [32], it is clear that:

$$\Delta\varphi = \mathrm{FT}^{-1}\left\{\frac{1}{3}\hat{\mathrm{H}}^{\mathrm{T}} \cdot \mathrm{FT}\{\chi\} \cdot \hat{\mathrm{H}} - \hat{\mathrm{H}} \cdot \vec{\mathrm{k}} \frac{\vec{\mathrm{k}}^{\mathrm{T}} \cdot \mathrm{FT}\{\chi\} \cdot \hat{\mathrm{H}}}{k^2}\right\}\gamma Ht \tag{0.57}$$

Equation 0.57 should be inverted to determine $\chi$. This is a well-known ill-posed problem because of the existence of zero coefficients in the right-hand side of equation 0.57. This difficulty goes away if we apply numerical regularization [47, 50-52] or acquire a set of phase images at different orientations with respect to the main field [47, 53]. We define a matrix A which contain n independent measurements in the subject frame of reference, and then $\chi$ can be described as below

$$\vec{\theta} = \left[\frac{\Delta\varphi_1(\vec{k})}{\gamma Ht} \quad \frac{\Delta\varphi_2(\vec{k})}{\gamma Ht} \quad \cdots \quad \frac{\Delta\varphi_n(\vec{k})}{\gamma Ht}\right]^{\mathrm{T}} \tag{0.58}$$

$$\vec{\chi} = \left[\chi_{11}(\vec{k}) \quad \chi_{12}(\vec{k}) \quad \chi_{13}(\vec{k}) \quad \chi_{22}(\vec{k}) \quad \chi_{23}(\vec{k}) \quad \chi_{33}(\vec{k})\right]^{\mathrm{T}} \tag{0.59}$$

$$\vec{\mathrm{A}} = \begin{bmatrix} \frac{1}{3}\hat{H}_1^{(1)}\hat{H}_1^{(1)} - \vec{k}\cdot\hat{H}^{(1)}\frac{k_1\hat{H}_1^{(1)}}{k^2} & \frac{2}{3}\hat{H}_1^{(1)}\hat{H}_2^{(1)} - \vec{k}\cdot\hat{H}^{(1)}\frac{k_1\hat{H}_2^{(1)} + k_2\hat{H}_1^{(1)}}{k^2} & \cdots & \frac{1}{3}\hat{H}_3^{(1)}\hat{H}_3^{(1)} - \vec{k}\cdot\hat{H}^{(1)}\frac{k_3\hat{H}_3^{(1)}}{k^2} \\ \frac{1}{3}\hat{H}_1^{(2)}\hat{H}_1^{(2)} - \vec{k}\cdot\hat{H}^{(2)}\frac{k_1\hat{H}_1^{(2)}}{k^2} & \frac{2}{3}\hat{H}_1^{(2)}\hat{H}_2^{(2)} - \vec{k}\cdot\hat{H}^{(2)}\frac{k_1\hat{H}_2^{(2)} + k_2\hat{H}_1^{(2)}}{k^2} & \cdots & \frac{1}{3}\hat{H}_3^{(2)}\hat{H}_3^{(2)} - \vec{k}\cdot\hat{H}^{(2)}\frac{k_3\hat{H}_3^{(2)}}{k^2} \\ \cdots & \cdots & \cdots & \cdots \\ \frac{1}{3}\hat{H}_1^{(n)}\hat{H}_1^{(n)} - \vec{k}\cdot\hat{H}^{(n)}\frac{k_1\hat{H}_1^{(n)}}{k^2} & \frac{2}{3}\hat{H}_1^{(n)}\hat{H}_2^{(n)} - \vec{k}\cdot\hat{H}^{(n)}\frac{k_1\hat{H}_2^{(n)} + k_2\hat{H}_1^{(n)}}{k^2} & \cdots & \frac{1}{3}\hat{H}_3^{(n)}\hat{H}_3^{(n)} - \vec{k}\cdot\hat{H}^{(n)}\frac{k_3\hat{H}_3^{(n)}}{k^2} \end{bmatrix} \tag{0.60}$$

$$\vec{\chi} = (\vec{\mathrm{A}}^{\mathrm{T}} \cdot \vec{\mathrm{A}})^{-1} \cdot \vec{\mathrm{A}}^{\mathrm{T}} \cdot \vec{\theta} \tag{0.61}$$

In order to define rotational invariant quantities after calculating out the tensor, we apply eigenvalue decomposition to the measured tensor and obtain three principal susceptibilities value [47]. They are denoted as $\chi_1$, $\chi_2$ and $\chi_3$ with a corresponding eigenvector. The principal eigenvector corresponds to the largest eigenvalue which exhibits the largest magnetic susceptibility. We further borrow the color-coding scheme from DTI and STI and apply it to the major principal susceptibility $\chi_1$ as follows: red represents anterior–posterior direction, green represents left–right, and blue represents dorsal–ventral, and then we put all three images together to achieve susceptibility anisotropy mapping.

# 4. Anisotropy Mapping in Biological Tissues

Rat brain images and trabecular bone [54] images are acquired using iMQCs between spins that are 10 μm to 500 μm apart. Because the dipolar interaction between spins is dependent on the correlation gradient direction [55], iMQC images in different directions can be used for anisotropy mapping. We investigated tissue microstructure by analyzing anisotropy mapping. At the same time, we simulated images expected from rat brain without microstructure. We compare those with experimental results to prove that the dipolar field from the overall shape only has small contributions to the experimental iMQC signal.

Besides magnitude of iMQCs, phase of iMQCs should be studied as well. The phase anisotropy maps built by the iMQC method can clearly show susceptibility information in rat brain. The susceptibility data may provide meaningful diagnostic information. To deeply study susceptibility, the modified-crazed sequence is developed. Combining phase data of modified-CRAZED images and phase data of iMQCs images is may allow one to construct microstructure maps. Obviously, the phase images in all above techniques need to have high-contrasted and need to be clear. To achieve that goal, the mathematics tools from Susceptibility-Weighted Imaging (SWI) and Susceptibility Tensor Imaging (SWI) are used for iMQC experiments.

In addition to rat brain imaging, we also investigate the use of iMQC to probe structural anisotropy in trabecular bone. Chung et al. [55] showed that T2* of the marrow surrounded by trabecular bone decreases from $\alpha$ = 0 degree to $\alpha$ = 90 degree, where $\alpha$ is the angle between the magnetic field and the long bone axis. Instead of changing $\alpha$ which is physically complicated in practice, the CRAZED sequence allows one to change the correlation gradient while leaving the sample stationary. The correlation gradient direction is parameterized by the parameter theta, where theta is the angle between $B_0$ and the correlation gradient. Below, it is shown that the signal vs. theta curves are different for a vegetable oil emulsion and red bone marrow, suggesting that the CRAZED sequence is sensitive to anisotropy in the trabecular bone.

## 4.1 Why iMQC instead of DTI or STI?

Intermolecular Multiple Quantum Coherences (iMQCs) are unique in that they provide a fundamentally different source of anatomic and functional contrast as compared to conventional MRI. iMQCs have been shown to non-invasively probe material microstructure in liquid state NMR [25]. iMQCs contrast comes from pairs of spins separated by a well-defined and user-selectable correlation distance. Typically, only distances between 10 μm and 500 μm can be probed. This distance scale is what we refer to as mesoscopic scale. Usually, the resolution is limited by the available magnetic field gradient strength (spins are resolvable if the gradient separates their frequencies by more than the intrinsic linewidth), but in practice, the inherent low sensitivity and limited scan time (particularly in vivo) normally provides the more fundamental limitation. As a result, the resolution of conventional clinical MRI images is limited to voxels much larger than cellular dimensions (on the millimeter scale, typical larger than 500 μm). There is another technique called Diffusion Tensor Imaging (DTI) that probes sub-voxels effects. DTI typically probes effects smaller than 10 μm. This is because, in bulk water, molecules diffuse isotropically, with root mean square motion of approximately 7 μm in any specific direction over 10 ms. In tissues of rat brains diffusion is anisotropic, giving access to local structure on the micrometer scale (usually smaller than 10 μm). However, intermediate regimes, where the length scale ranges from around 10 μm to 500 μm are still generally difficult to access. So, being able to probe features on this mesoscopic scale makes iMQCs unique.

In this sense, iMQCs are important in that many examples of porous materials in vivo have structures on the mesoscopic scale [56]. For example, Trabecular bone consists essentially of an array of interconnected struts typically in the mesoscopic scales which form a structurally anisotropic network [57]. When bone loss, which occurs in postmenopausal osteoporosis or extended exposure to microgravity happens [12], the structure and hence the degree of anisotropy and topology changes. Bone loss is associated with the progress of disease and monitoring the effects and progress of novel therapies. Conventional clinical MRI methods cannot spatially resolve trabecular bone structures; DTI is also not an option because the bone pores are too large. When the material inclusions or pores are large compared to the mean diffusion length of water, it will be very time-consuming and ineffective use DTI. iMQCs, on the other hand, stand out as a good method because they encode material geometry on a mesoscopic length scale.

iMQCs and DTI detect anisotropy in different ways. In DTI, the anisotropy of water diffusion shows us sub-resolution brain structure, which facilitates diagnosis and grading of malignant brain tumors, delineation of white matter fiber tracks and even analysis of Major Depressive Disorder. Anisotropy can be defined as a directional dependence in a material's physical properties, including conductivity and susceptibility. Traditionally, anisotropy is produced by Diffusion Tensor Imaging (DTI). DTI measures the fractional anisotropy of the random motion (Brownian motion) of water molecules. Water will diffuse more rapidly in the direction aligned with the internal structure, and more slowly as it moves perpendicular to the preferred direction. This causes the anisotropy.

In contrast, iMQCs detect anisotropy by a different mechanism, although the precise mechanism is unknown. From previous research, the iMQCs anisotropy of trabecular bone indicates the present condition of bone [56]. It is highly likely that the fractional iMQCs anisotropy in the brains can be exploited to create a map of the fiber tracts. So, anisotropy mapping is a very interesting and powerful application for iMQCs.

Susceptibility Weighted Imaging (SWI) [32] and Susceptibility Tensor Imaging (STI) [30, 31, 47] can also detects anisotropy in tissue. The magnetic susceptibility leads to a resonance frequency shift which can be measured using a multi-gradient-echo sequence. In 1987, Young et al. [58] used phase maps to detect changes in the local magnetic field in tumors, lacunar infarct, and multiple sclerosis. They attributed the effect to the paramagnetic contributions of species such as deoxyhemoglobin, methemoglobin, free ferric iron, hemosiderin, and other breakdown products of blood [58]. However, in most situations the phase maps are discarded because of its poor contrast. Later, Haacke et al. found that if combine the phase image with the magnitude image, enhanced contrast image focus on the scale of around 10 microns can be obtained [30, 31]. This is a very useful tool to see venous blood, hemorrhage, and iron storage [59, 60]. In detail, it needs phase unwrapping and background noise removal which includes going through high pass (HP) filter, sphere polynomial fitting, mean value filtering followed by a de-convolution and projection onto dipole field [38, 43, 46, 61].

Based on Susceptibility Weighted Imaging, Susceptibility Tensor Imaging (STI) [47], has been proposed and proved showing distinctive fiber pathways in 3D in the mouse brain [62]. Generally, it has been assumed that susceptibility is isotropic (orientation independent) in biologic tissues [47]. It is similar to diffusion in this perspective. Liu et al. observed the orientation-dependent magnetic susceptibility in the mouse central nervous system, proposed a method to show apparent susceptibility tensor and proved the susceptibility anisotropy can be an intrinsic property of tissue or can be purposely induced by the introduction of external molecular agents [31, 58].

However, the STI has a huge limitation to measure the susceptibility anisotropy. It requires the measurement of susceptibility at different orientations by rotating the sample with respect to the main magnetic field which is very difficult, time-consuming, and almost impossible in vivo. iMQCs, in contrast, can measure physics property in dependence with orientations by applying correlation gradients in different directions. This makes the experiments much easier to perform and possible in vivo. In addition to producing iMQC anisotropy by using crazed sequence, we also apply modifications to crazed sequence in order to obtain susceptibility anisotropy. We call it iMQC-susceptibility anisotropy. It uses crazed and mod-crazed sequence to detect mesoscale anisotropy without moving the sample. iMQC-anisotropy measurements also benefit from the mathematics tools to make susceptibility information or phase information clear and highly-contrasted. We found that the iMQC phase data correlate much better with rat kidney anatomy as compared to magnitude data. This improved correlation can unveil some underlying physical effects related to the resonance frequency offset, susceptibility, etc. They can be used to create images that can be co-registered with those methods and anatomic images, in order to highlight what is new and different.

In addition, I also will briefly introduce the dipolar field treatments which discuss the fundamental source of iMQCs in liquids.  A set of experiments will then be discussed producing optimized intermolecular Double Quantum Coherence (iDQCs) images. I will also discuss the simulation of iDQCs signals in isotropic media. Thereafter, I explain the way to construct anisotropy maps by combining iDQC images.  iMQC anisotropy maps are acquired in ex vivo kidney. In the future, they can be used to create images that can be co-registered with those other methods and anatomic images, in order to highlight what is new and different.

## 4.2 Methods and samples

The NMR sequence we use is COSY Revamped by Asymmetric Z-gradient Echo Detection (CRAZED) which was first introduced in the early 90s and showed strong unexpected iMQCs signals [9] that can only be explained when including dipolar coupled spin pairs.  The principle of iMQCs is explained using the high temperature approximation and DDF framework.

One of our samples, rat brain, is scanned at 7T.  The brain is doped with gadolinium (prohance). The correlation distance was 70 μm (correlation distance is explained later). The 90 degree and 120 degree pulses are Gaussian pulses. Adiabatic hyperbolic secant pulsed were used for two refocusing 180 degree pulses. The repetition time for all images was 1 s which is much larger than T1 of rat kidney which was measured to be 40ms. The slice thickness was 2 mm with a 3cm*3cm FOV. The acquisition matrix size was 256*256.

Before taking iDQC scans, spin echo images were also acquired. The 90 degree and 180 degree pulses were Gaussian pulses. The repetition time for all images was 14 s. The echo time was 10 ms for rat kidney. The slice thickness was 2 mm with a 3cm*3cm field of view. The spin echo acquisition matrix size is 256*256.

In addition, we scanned the trabecular bone sample with $\alpha$ changing ($\alpha$ is the angle between the magnetic field and the correlation gradient direction). For better and necessary understanding, we also performed control experiments on vegetable oil emulsion. The data analysis proves credibility of red bone marrow showing anisotropy instead of noise.

## 4.3 iDQC-Crazed simulations

The signal in a typical Crazed sequence [17, 64] is approximately equal to the normal magnetization (proportional to proton density) multiplied by the dipolar field. This formula can be explained from:

$$\frac{d\overrightarrow{M}(\vec{r},t)}{dt}=\gamma\overrightarrow{M}(\vec{r},t)\times[B_0\hat{z}+\overrightarrow{B}_d(\vec{r},t)] \tag{0.62}$$

Specifically, $\vec{M}(\vec{r},t)$ on the right of equation (0.62) can be calculated by applying a modulation to the magnetization signal in the frequency domain. The modulation here means a certain correlation distance $d_c$ in frequency domain determined by:

$$d_c = \frac{\pi}{\gamma GT} \tag{0.63}$$

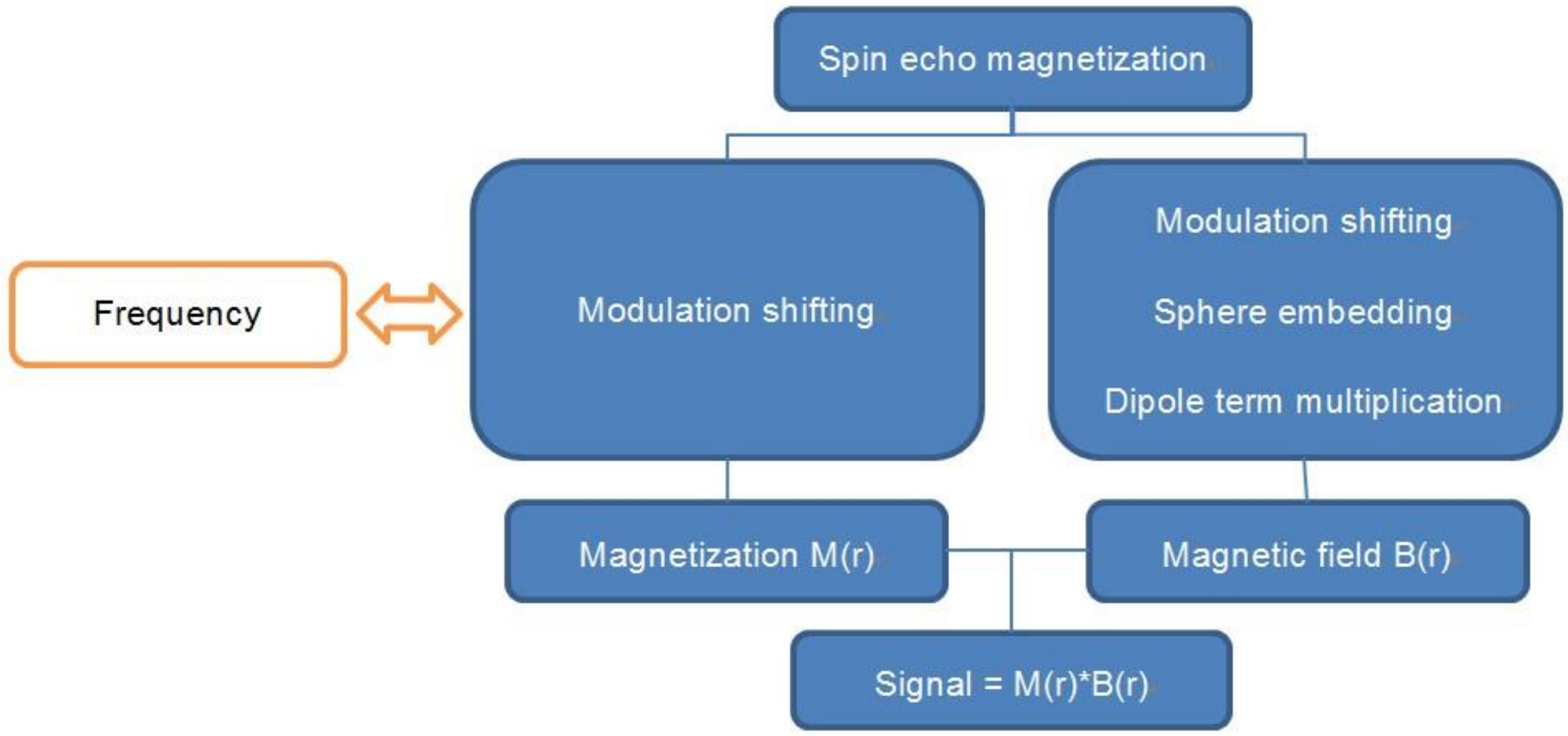

**Figure 8: The process of iDQC simulations: Input the spin echo magnetization, perform modulation in the k space including uniform dipole term multiplication, Fourier transform back those images and construct simulated iMQCs.**

$B_d(\vec{r})$, which is the dipolar field, depends on the magnetization distribution and thus on the shape of the object [63, 65]. The uniform dipolar field can be found by Fourier transformation of the magnetization density, multiplying by the uniform dipolar term $(3(\hat{k}\cdot\hat{z})^2\text{-}1)/2$ [28], followed by an inverse Fourier transform. This can easily be performed for any shape given a high resolution image that provides the magnetization distribution. However, before multiplying the dipolar term, we should apply the same modulation to $\vec{M}(\vec{r},t)$, and then embedding a sphere. The step "Sphere embedding" is not only for standardization of the frequency domain images for uniform comparisons between images, but also for transferring the information of the center point to other points without causing any distortions to the dipolar field. The center point of dipole term in frequency domain is a singular point. Its value becomes 0 by embedding a sphere symmetrically, at the same time the singularity goes to infinite location

## 4.4 iDQC anisotropy mapping

Now that we understand the physical basis of iMQCs, it is interesting to use iMQCs to construct anisotropy images. Previous work has shown that iMQC-based anisotropy measurements can detect anisotropy in structured materials (such as tumors with embedded nanoparticles) [12, 56, 66].

For typical imaging applications, iMQC images are more sensitive to the susceptibility interfaces than spin echo images. By varying the direction and amplitude of the correlation gradient pairs, unique contrast can be obtained. The direction of the correlation gradient pair affects the contrast because the dipolar field is proportional to $(3\cos^2\theta\text{-}1)/2$, where $\theta$ is the angle between $B_0$ and the direction of the correlation gradient. As a result, the signal for a Z correlation gradient should be the opposite sign and two times larger than the signals for X and Y correlation gradients. In isotropic media, adding the three complex images together or subtracting the magnitude images should yield 0. On the other hand, in anisotropic media, these combinations reflect the local structure. The strength of the correlation gradients determines the distance between the two coupled spins. In isotropic media, the correlation distance should have no effect on the signal (ignoring the diffusion and sample size limitations). For anisotropic media, however, the correlation distance affects the image contrast because the signal is directly related to the dipolar field at that distance.

The iDQC signal formula depends on many factors including [64] the magnitude and direction of the dipolar field, $T_2$, the resonance frequency offset, the dipolar demagnetizing time ( $\tau_d = (\gamma \mu_0 M_0)^{-1}$ ), pulse flip angles, etc. But roughly, |Z+Y+X| and |Z|-|Y|-|X| should be 0 for isotropic areas and show signals for anisotropic areas. For the simulation, generally only uniform dipolar field is considered which means that |Z+Y+X| and |Z|-|Y|-|X| should be very close to 0. Previous studies have shown that mesoscopic structural anisotropy maps can be obtained with iMQCs [12, 56] [67-69].

## 4.5 iDQC-susceptibility imaging & iDQC-susceptibility anisotropy mapping

Besides using iMQC magnitude data to construct anisotropy images, we are still trying to find other ways or improvements to construct anisotropy and explain its physics basis. After several new experiments, we found that fractional phase anisotropy maps are very clear and informative. Fractional phase anisotropy maps are produced by following steps: take the phase data of the iMQC signals and choose a small area of rat kidney which is very flat and isotropic, and then match the values of that area in Z image, Y image and X image to be 2:-1:-1 by multiplying the three images by constants. Multiply those constants to every pixel value of their related image, respectively. At last, construct the anisotropy maps using previously explained methods.

At the first glance this method seems like having no physical meaning. However, considering that the iDQC phase data actually contains susceptibility information, it is very likely that the phase data intensity in dependence with correlation distance orientations can be clearly studied. More importantly, the results truly show us some useful information.

To dig more deeply, consider the effect of changing the iDQC sequence in the following way. The Standard-Crazed signals have contrast from both magnetization density and resonance frequency variation; so-called modified-CRAZED sequences, with an extra 180 degree pulse in the middle of Tau interval, can produce images which only have contrast from magnetization density variation. So the difference between Standard-Crazed signals and Modified-Crazed signals from the XY terms of the dipolar field predicts local phase shifts from susceptibility or magnetization anisotropy. This has prompted a more careful examination, borrowing the tools recently developed for phase interpretation in susceptibility weighted imaging. The raw phase from such images is dominated by unwanted artifacts such as coil phase shifts and shim effects. Thus the observed phase requires several corrections (phase unwrapping, fitting and deconvolution) and projection onto a dipole field which assumes susceptibility proportional to magnetization. We call this method iMQC-susceptibility imaging. If we repeat iMQC-susceptibility imaging for X, Y and Z terms, and combine them using techniques form Susceptibility Tensor Imaging, we obtain iMQC-susceptibility anisotropy mapping.

## 4.6 Results and discussions

### 4.6.1 Rat brain imaging

A growing number of researchers are putting focus on rat brain. In fact, the brains of rats are strikingly similar to those of humans. Extensive research in diseases of the brain, such as Alzheimer's, is usually conducted by means of rat models. Thus, use of rat models accelerates our understanding of human disease and behavior. For instance, recent papers [70] published by Jared Smith and Kevin Alloway detail their discovery of a parallel between the motor cortices of rats and humans that signifies a greater relevance of the rat model to studies of the human brain than scientists had previously known.

Two brain structures exhibited anisotropy in iMQC images (results shown below), and those two structures will be described here. First, the corpus callosum (CC) is a wide, flat bundle of neural fibers beneath the cortex in the eutherian brain at the longitudinal fissure. It connects the left and right cerebral hemispheres and facilitates inter-hemispheric communication. In the adult rat brain, diffusion in the somatosensory cortex is isotropic, while in the corpus callosum [71, 72] and hippocampus it is anisotropic [73]. In human brain, mean diffusivity is significantly increased and fractional anisotropy is significantly reduced in the splenium (posterior of CC) but not the genu (anterior of CC) of the corpus callosum in the schizophrenic group compared with controls [74]. The size of the splenium of the corpus callosum can also tell the sex differences in human and rats [75]. The older alcoholics have smaller genu and splenium and higher diffusivity in these regions than younger alcoholics [76]. Significant alterations are revealed in the molecular diffusion and in the size of the CC with respect to gender and handedness [77]. Even the growth time of CC is studied [78, 79]. Thus, studying corpus callosum is very necessary.

Optic chiasm (OC) is the part of the brain where the optic nerves partially cross. It is located at the bottom of the brain immediately below the hypothalamus. Classically, it has been known as the place where groups of retinal axons segregate to pass into the optic tract on either the same or the opposite side of the brain. This segregation of the axons into a crossed and uncrossed component allows the appropriate bilateral connections that underlie normal binocular vision to form [80]. The importance of understanding architecture and development of optic chiasm has been stressed widely [81-85].

Corpus callosum and optic chiasm are clearly shown in iDQC experiments results (discussed later). This provides a whole new approach to the structure and development of CC and OC, which highly likely gives us information about rat or even human's behavior and diseases.

Before directly showing the anisotropy maps, we would like at first to compare the intermolecular multiple quantum coherences images with different kinds of other conventional MRI images (Fig. 9). Coronal section images of rat brain are shown here. Clearly, the contrast rising from iDQC is more obvious than any other conventional MRI shown here. Corpus callosum and optic chiasm are very clearly shown and have a very nice contrast. More important, unlike conventional MRI, where image contrast is largely based on variations in spin density and relaxation times (often with injected contrast agents), contrast with iDQC images comes from dipolar couplings in intermediate scales dictated by gradient strength. In the rapidly expanding field of functional MRI, contrast is frequently the limiting factor. New methods for contrast enhancement could thus improve tissue characterization, particularly if they correlate with physiologically important characteristics. It has already been reported that this contrast is useful in the detection of small tumors [67] and in functional MRI [12, 56].

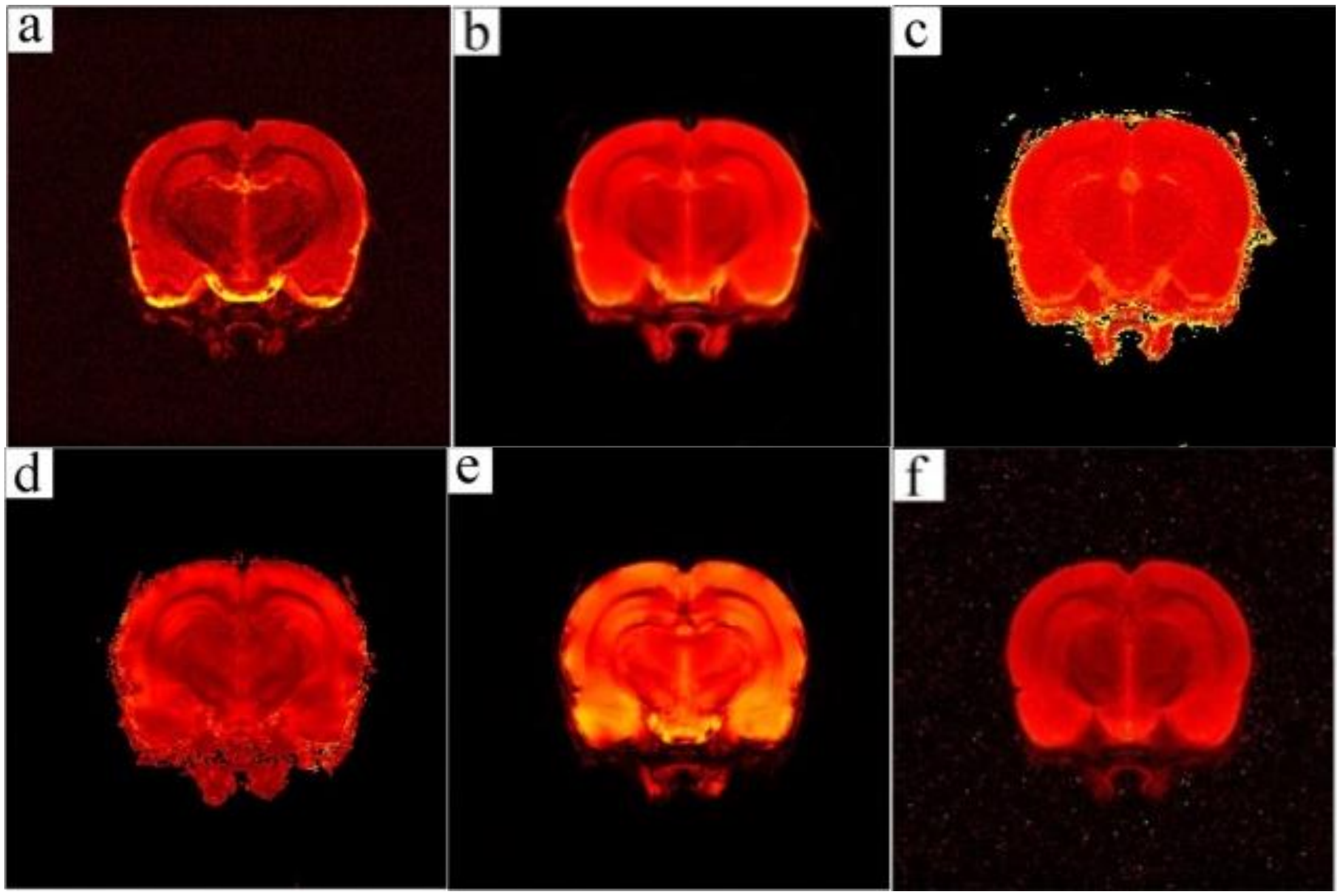

**Figure 9: All the images are rat brain images scanned at 7T, the scanning thickness is 2mm, the field of view is 2.5cm*2.5cm. (a). iDQC-Crazed image (b). Spin echo image (c). T2 map (d). T2* map (e). Proton density map (f). Diffusion trace weighted image (It is calculated as s0*exp(-b*trace), s0 is the diffusion experiment signal without gradient, b is a constant related to diffusion gradient, The trace is calculated after diagonalization to the diffusion tensor)**

Now we can consider iMQC-based anisotropy mapping. Usually MRI sequence contains 3 channels which are slice selection channel, phase encoding channel and frequency encoding channel. For the iDQC sequence introduced above, we can put the double quantum filter (gradients GT, 120 degree pulse and 2GT) in different channels which are, the slice selection channel (physical Z direction), the phase encoding channel (physical Y direction) and the frequency encoding channel (physical X direction). Thus, iDQC images in different physical directions are obtained (top images of Fig. 10). From the general intensity and color scales, it is easy to confirm that the intensity of the Z-crazed image is roughly twice that of Y-crazed or X-crazed, which proves that the experiments roughly accords with the dipolar field term on spin echo image (manually dipolar field effects is added). The bottom images of Fig (10) show the anisotropy. The $|Z+Y+X|/\max(|Z|)$ map contains phase information while $(|Z|-|Y|-|X|)/\max(|Z|)$ doesn't. The anisotropy map without phase information seems to have larger contrast. Currently the clinical information which can be extracted is that iDQC anisotropy maps clearly show us the shape of corpus callosum and optic chiasm. Future research topics can be the correlation between the shape or size of CC or OC and age or specific behavior. It will be interesting in future work to test if iMQC anisotropy measurements correlate with pathological conditions.

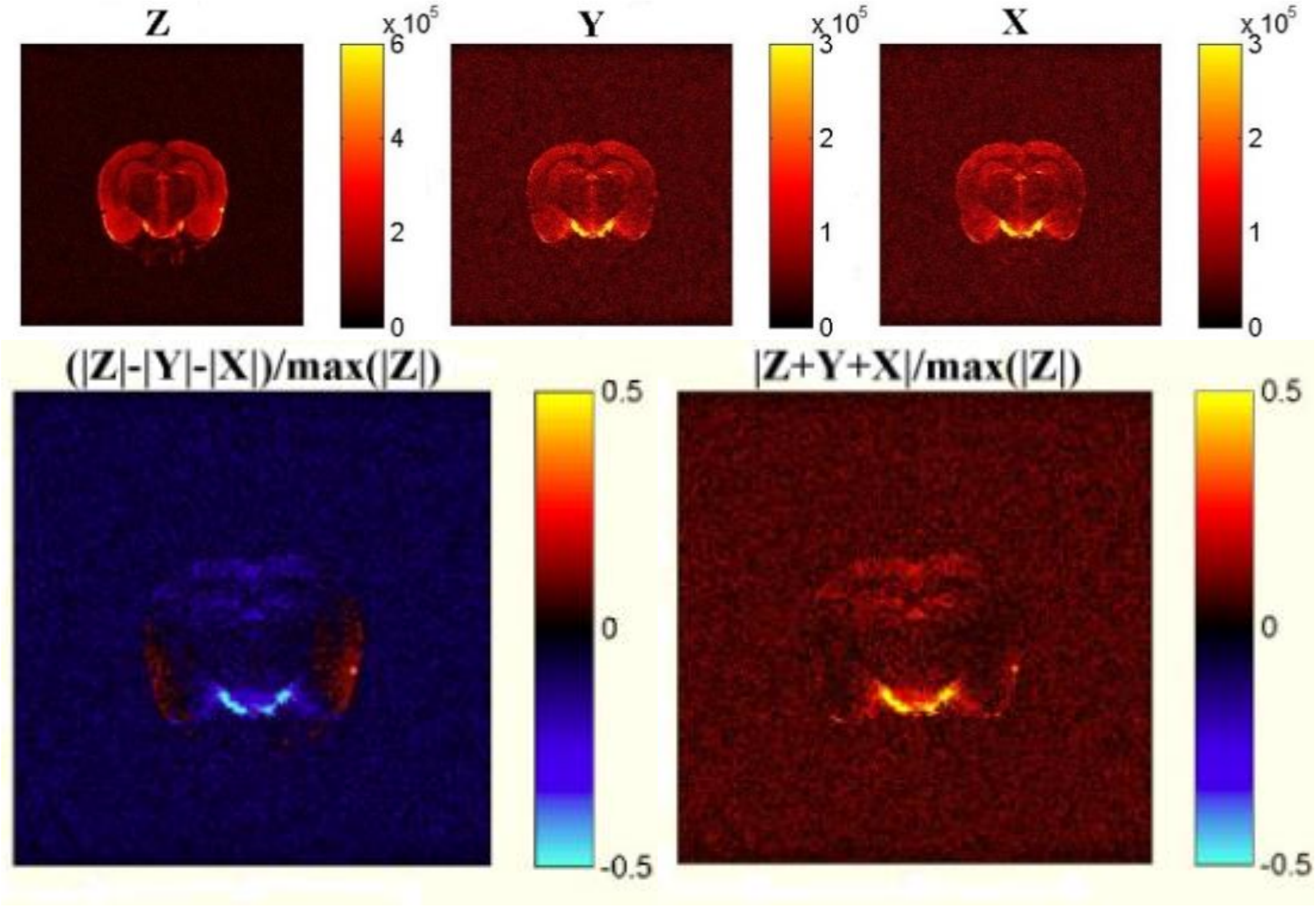

**Figure 10: iDQC-Crazed rat brain experimental images. The main magnetic field direction is along Z direction. The top row is the intensity maps produced by considering correlation distance in 3 directions Z, Y and X. The bottom row displays the fractional anisotropy maps which are calculated by (|Z|-|Y|-|X|)/max (|Z|) and |Z+Y+X|/max (|Z|).**

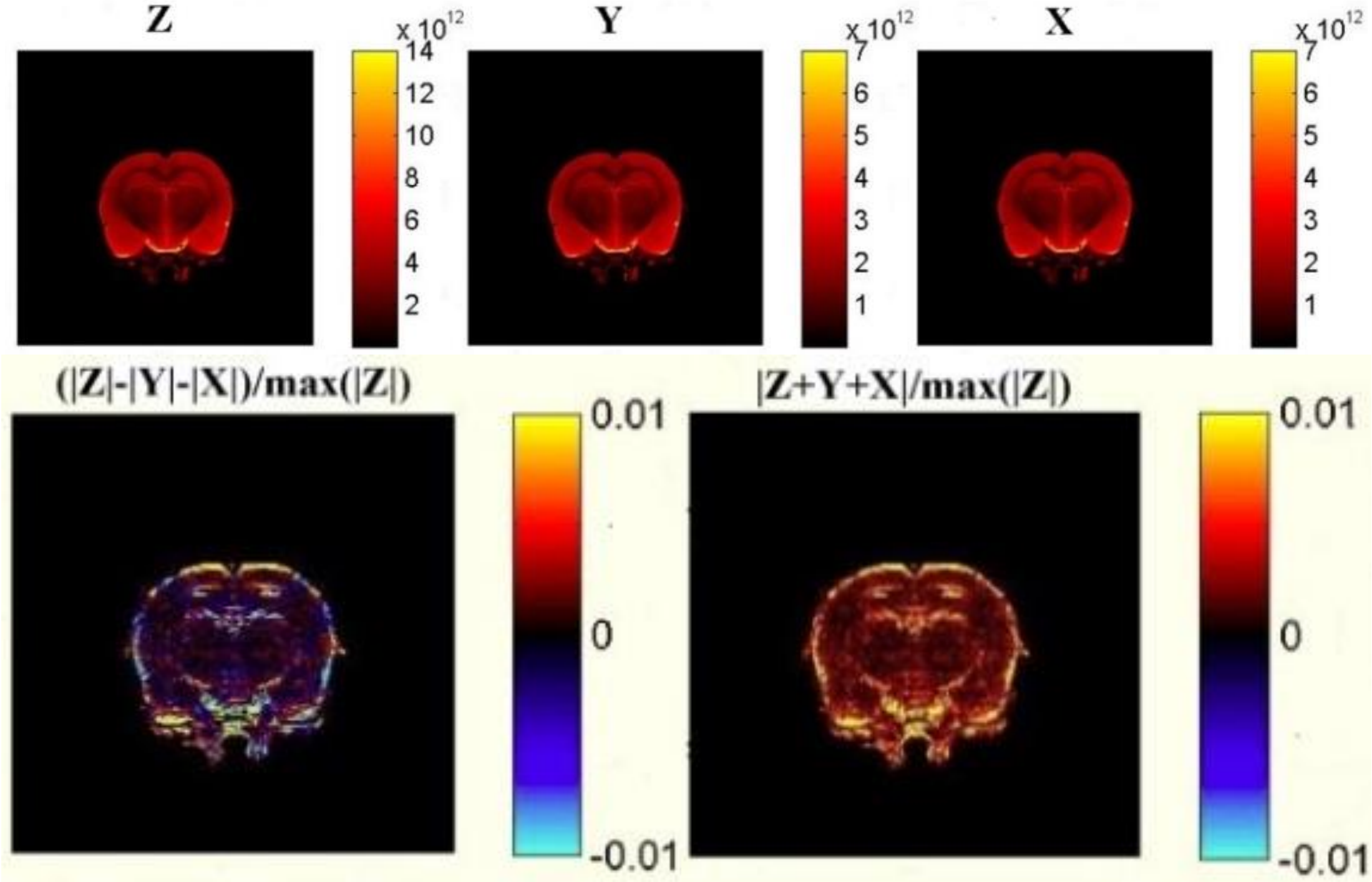

**Figure 11: iDQC-Crazed rat brain simulated images. The top row displays the simulated images which are produced applying correlation distance to spin echo density (S) weighted map in 3 directions, Z, Y and X. The bottom row displays the fractional anisotropy maps (|Z|-|Y|-|X|)/max (|Z|) and |Z+Y+X|/max (|Z|) a by uniform dipole field term.**

Fig (11) shows the simulated fractional magnitude anisotropy images, (|Z|-|Y|-|X|)/max (|Z|) and |Z+Y+X|/max (|Z|). These simulations prove that the theory of iDQC images and corresponding anisotropy measurements are correct and could be very powerful. The intensity of Z-Crazed is roughly twice of that of Y-Crazed and X-Crazed, which is consistent with the directional dependence of the dipolar field ($(3\cos^2\theta\text{-}1)/2$). More quantitatively, one can calculate (|Z|-|Y|-|X|)/max (|Z|) and |Z+Y+X|/max (|Z|) using simulated data for the anisotropy maps, both of which should

be zero in isotropic tissues. The experimental results show that the above quantities are non-zero, suggesting that there is anisotropy in those brain regions. In the case of simulation, theory shows that we should see no anisotropy for uniform dipole field situation. In reality, experiments show tissue microstructures in the rat brain in the bottom images of Fig (11). Is it wrong? Actually, this is just a trick of choosing different color-bar. In Fig (12), I choose the same color-bar for experimental and simulated images which is easier to compare. Clearly the conclusion is that no fractional anisotropy is detected for uniform dipolar field, which means the simulation we did is correct.

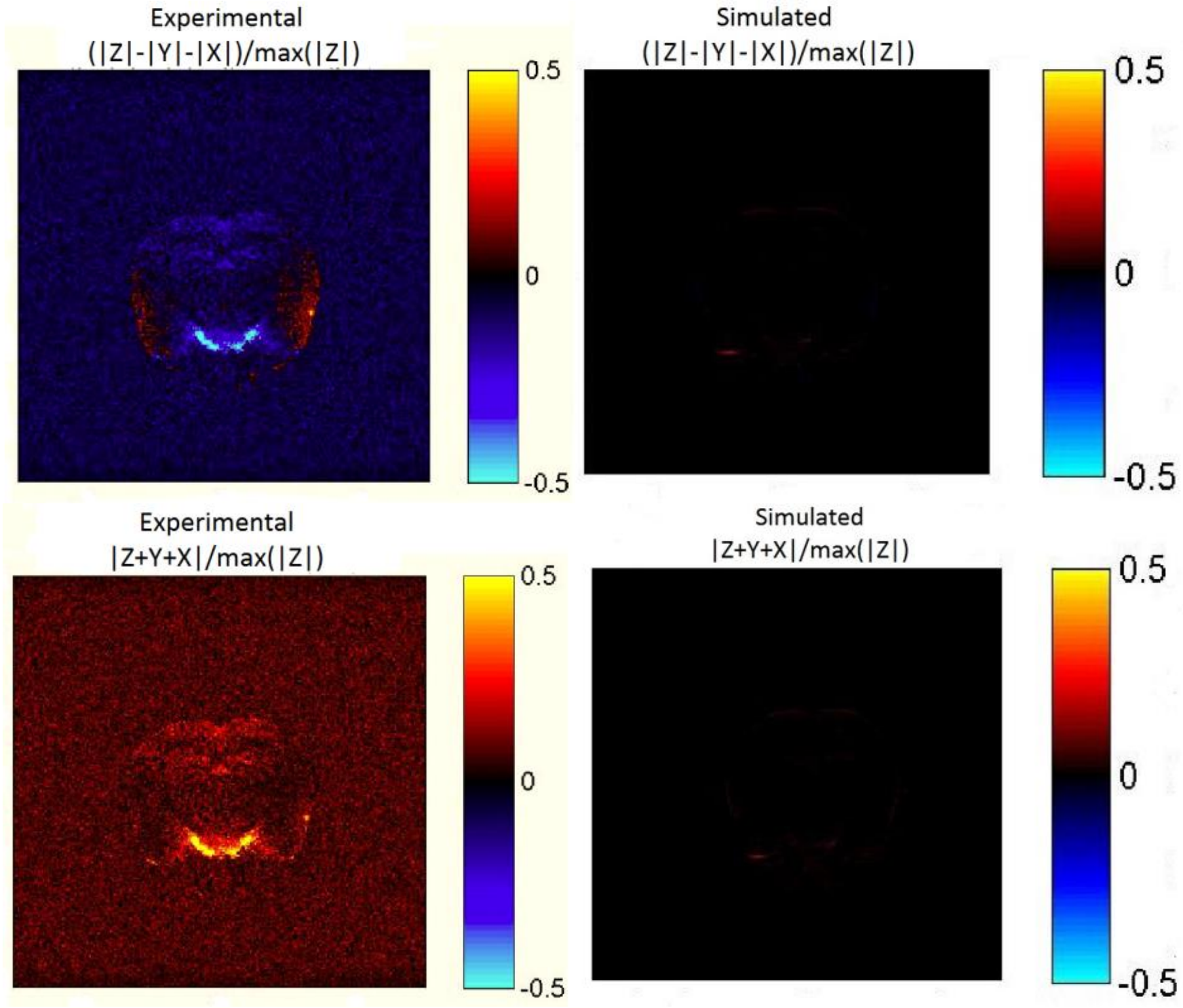

**Figure 12: Experimental fractional anisotropy maps and simulated fractional anisotropy maps in the same color-bar. The first and third images are experimental**

**(|Z|-|Y|-|X|)/max (|Z|) map and |Z+Y+X|/max (|Z|) map. The second and fourth images are simulated (|Z|-|Y|-|X|)/max (|Z|) map and |Z+Y+X|/max (|Z|) map.**

It also drops a hint that this perspective is very promising when comparing the T2 map and T2* map in fig (9). Clearly, their contrast is very different which means that the magnetic field inhomogeneities and susceptibility effects play an essential role in T2* map. So digging into the susceptibility-related images makes perfect sense.

We applied standard crazed sequence and mod-crazed sequence to rat brain. We acquired Fig. 14. Obviously it's seen that the optic chiasm has different signal intensity in those images which indicates some susceptibility information.

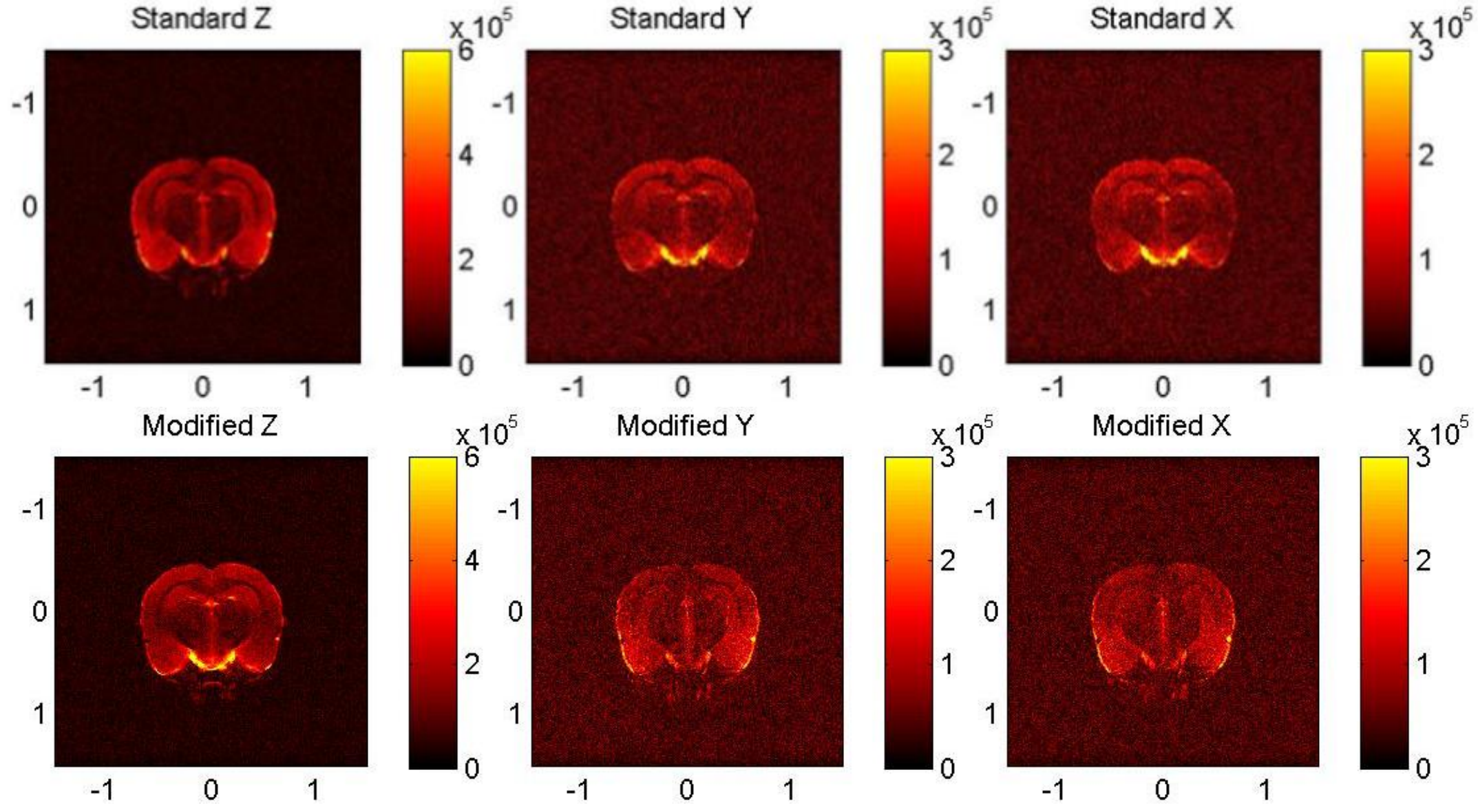

**Figure 13: Standard-Crazed and Modified-Crazed rat brain images. The top row is the intensity maps produced by applying correlation distance in 3 directions, Z, Y and X. The bottom row is the corresponding modified maps which insert 180 degree pulse in the middle of Tau interval. You can clearly see the different contrasts**

**between these two sets of images. Notice that standard crazed sequences and modified crazed sequences have the same timing after 120 mixing pulse.**

Importantly, iMQC-susceptibility anisotropy mapping is a new method and these phase-relevant results show a total new contrast which can be co-registered with many other MR methods, and it gains a promising expectation in clinical research because they are looking at the proper scales 10 μm to 500 μm. Considering what is said in the introduction part that, in anisotropy mapping, iDQC images have huge advantages over susceptibility tensor imaging technique, we should really use iDQC effects to do anisotropy-related practical research. Actually what susceptibility-related images can measure, iDQC images can measure that, too. For examples, brain iron concentrations in vivo [86] and unprecedented anatomical contrast in both white and gray matter regions [61, 62, 87]. The clinical potential of susceptibility-related images is still under investigation but it is anticipated that it will provide novel insights into disease induced tissue change [88]. Thus, it is highly possible that intermolecular multiple quantum coherences images can also give us those insights without suffering the difficulty of susceptibility tensor imaging experiments.

### 4.6.2 Trabecular bone imaging

Prostate and breast cancers are two of the most common types of cancer in United States, and they metastasize to bone in more than two thirds of patients [89-91]. This process is named as red bone marrow metastases which are currently incurable [92, 93], even though radiation and bisphosphonates can temporarily relieve pain from bone fracture.

Many researchers are interested in using heat to treat bone metastases. Heat can allow targeted drug delivery to bone [94, 95], ablation of cancer cells in bone [96, 97], and palliation of bone pain [98-100]. Ryan M. Davis in Warren group also proposed using iZQCs to measure temperature in red bone marrow [89]. However, to detect these red bone marrow metastases in an early phase might be much more critical.

Andrea et al. [101] presented the recent results and technical developments of diffusion-weighted imaging on bone marrow and bone marrow pathologies. In the spine, DWI has proven to be a highly useful method for the differential diagnosis of benign and malignant compression fractures. In these pathologies, the microscopic structure of bone marrow is altered in a very different ways, leading to different water mobility, which can be depicted by DWI [101, 102].

However, these DWI-related sequences are frequently affected by artifacts, mostly caused by physiological motion. Therefore, the introduction of additional correction techniques, or even the development of new sequences is necessary [102]. iDQC anisotropy mapping stands out as an alternative method, and the results of experiments below show that it is a very promising and clear method which definitely should be developed more.

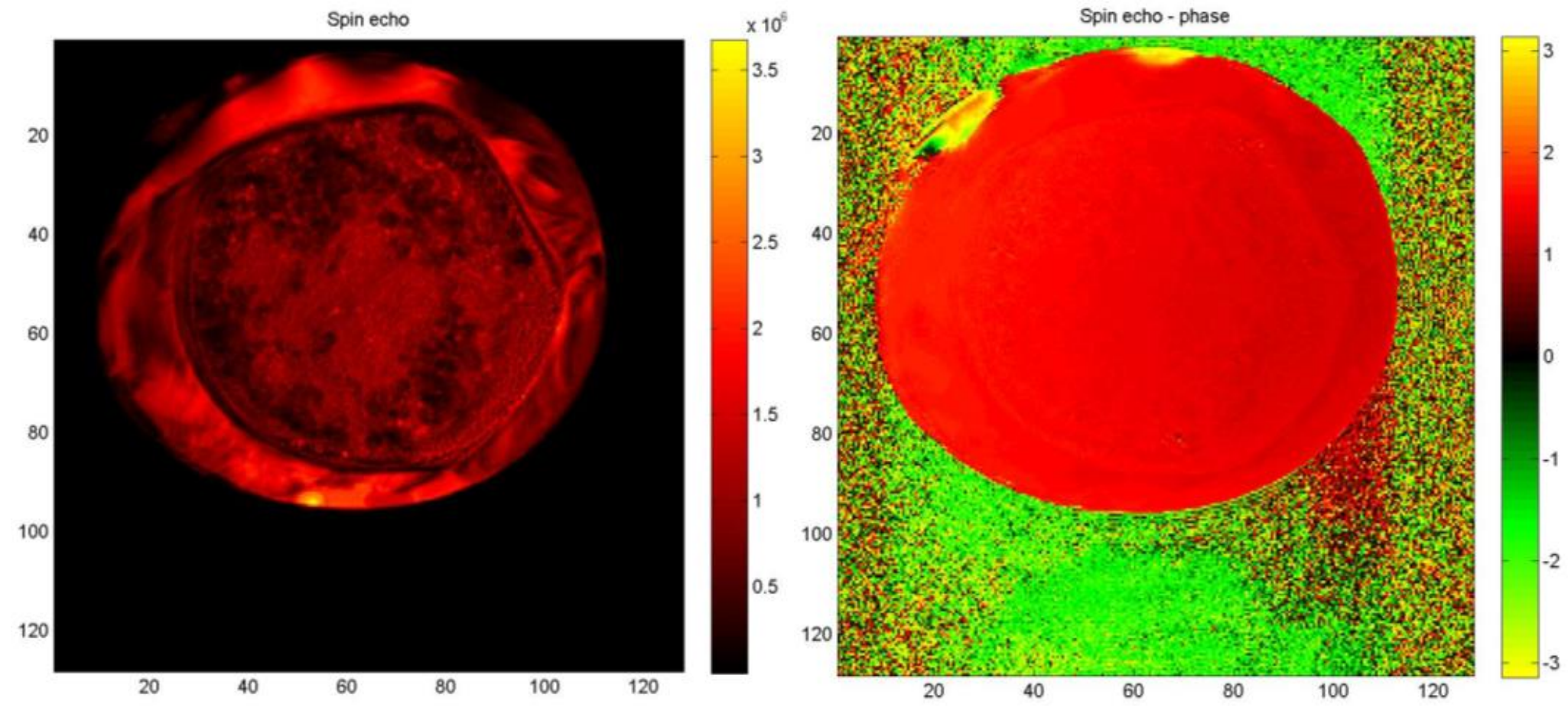

**Figure 14: The left imaging is the spin echo image of red bone marrow. The right imaging is the phase imaging of spin echo image of a red bone marrow.**

We are using porcine rib red marrow samples purchased from a grocery store. Soft tissue was removed around the bone using a scalpel, while still some muscle tissue left around the red bone marrow. Individual rib samples were frozen and saved for individual use. Figure 14 and 15 shows some basic iDQC images of red bone marrow.

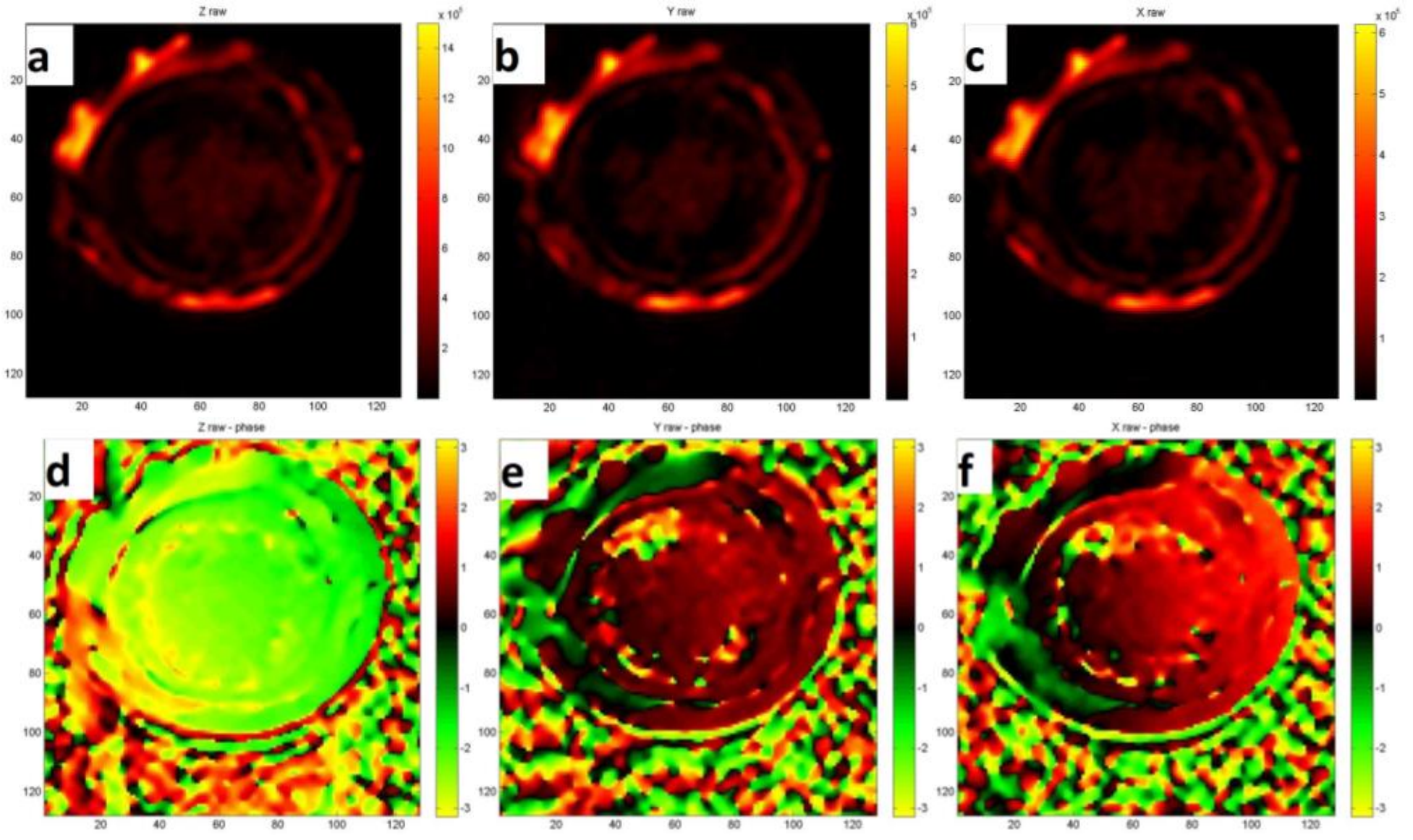

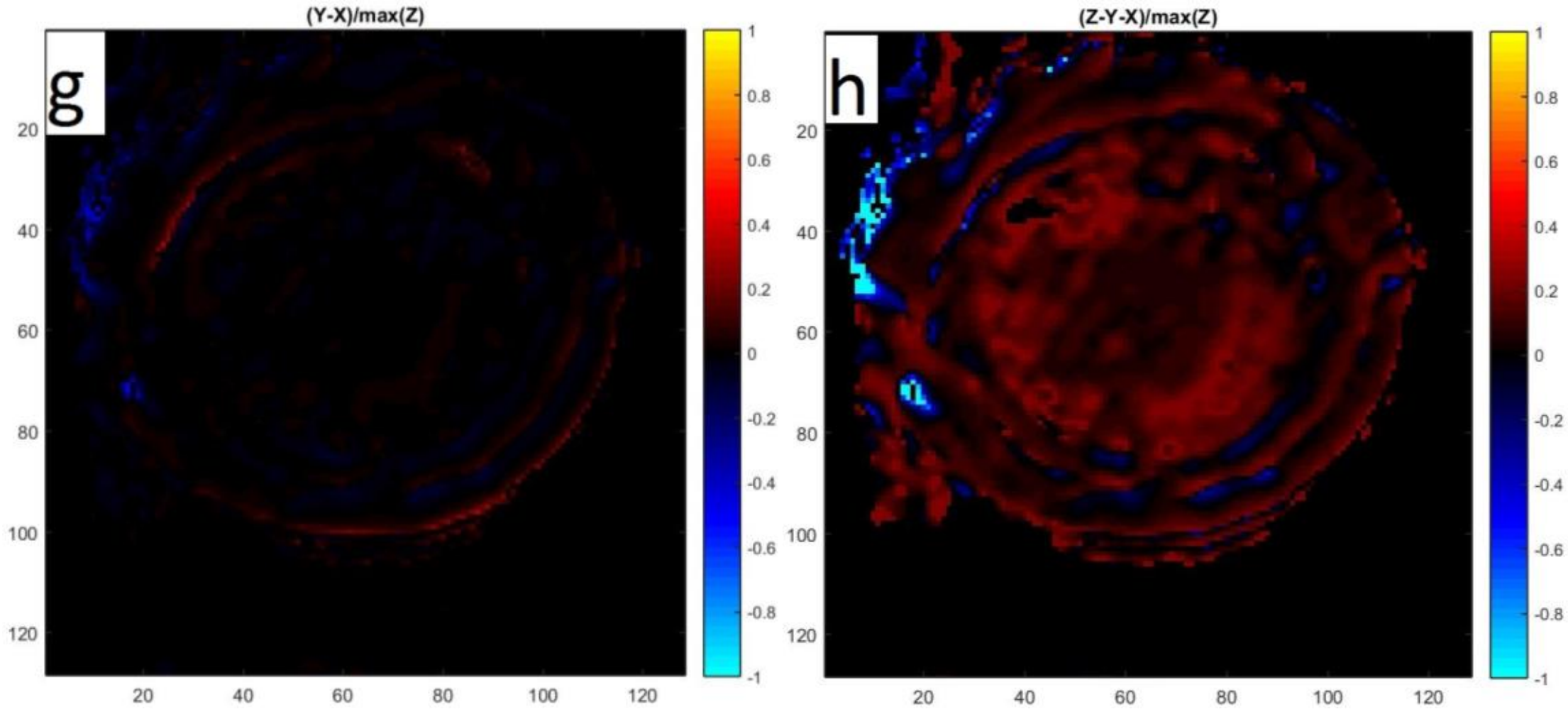

**Figure 15: (a), (b) and (c) are respectively magnitude images of double quantum coherence for Z axis, Y axis and X axis. (d), (e) and (f) are respectively phase images of double quantum coherence for Z axis, Y axis and X axis. (g) and (h) are processed images with magnitude of Z, Y and X images. (g) is (Y-X)/max(Z), and (h) is (Z-Y-X)/max(Z).**

The magnitude images for Z, Y and X look almost the same, while their intensity is different if you zoom in those images. The color-bar limit of Z is almost two times as that of Y and X, which indicates the rule $3\cos^2(\theta)-1$ for isotropic area.

Before producing images (g) and (h) in figure 15.B, a novel method to decrease noises in their magnitude images can be applied. Let's assume that noises everywhere can be described as $n_i e^{i\varphi_i}$, and the signal of the sample can be described as $S$ (real and positive). Considering Taylor expansion, $f(x) = f(a) + \frac{f'(a)}{1!}(x-a)$, we have,

$$
\begin{aligned}
mag &= \sqrt{\left(S + n_i \cos\varphi_i\right)^2 + \left(n_i \sin\varphi_i\right)^2} \\
&= |S| \cdot \sqrt{\left(1 + \frac{n_i}{S}\cos\varphi_i\right)^2 + \left(\frac{n_i}{S}\sin\varphi_i\right)^2} \\
&\approx |S| \cdot \left(1 + \frac{n_i}{S}\cos\varphi_i + \frac{n_i^2}{2S^2}\cos^2\varphi_i + \frac{n_i^2}{2S^2}\sin^2\varphi_i\right) \\
&= |S| \cdot \left(1 + \frac{n_i}{S}\cos\varphi_i + \frac{n_i^2}{2S^2}\right)
\end{aligned}
\tag{0.64}
$$

At first, $\frac{n_i}{S}\cos\varphi_i$ goes to zero when you apply averaging in $\varphi$. Then we go specific in 3 different directions, and we find out the relationship between signal and noise,

$$
\begin{aligned}
E_z &= S_z + \frac{n_i^2}{2S_z} \\
E_y &= S_y + \frac{n_i^2}{2S_y} \\
E_x &= S_x + \frac{n_i^2}{2S_x}
\end{aligned}
\tag{0.65}
$$

Equation (0.65) can be simply solved. For noise area, of course we set it up to be 0. For those signal area, the signal should be,

$$
\begin{aligned}
S_z &= \frac{E_z + \sqrt{E_z^2 - 2n_{iz}^2}}{2} \\
S_y &= \frac{E_y + \sqrt{E_y^2 - 2n_{iy}^2}}{2} \\
S_x &= \frac{E_x + \sqrt{E_x^2 - 2n_{ix}^2}}{2}
\end{aligned}
\tag{0.66}
$$

Practically root mean square $n_i$ is calculated from pure noise area. The image (g) basically shows us the difference between Y-crazed and X-crazed. Most areas in red bone marrow indicate identical except a small amount of pixels is different. However, (h) image, Z-Y-X image, clearly shows us the anisotropy information in the central parts of red bone marrow. The muscle tissues around it stay zero because they have no anisotropy.

We design a method to prove that it is truly anisotropy instead of just noise in the central part of red bone marrow. In figure 2, $\tau$ is the period that iDQC signals are produced. Longer $\tau$ , which is longer relaxation time, leads to smaller echo signals. We are comparing two situations of $\tau$ = 3.15 ms and $\tau$ = 4.15 ms. Of course we expect that former one has larger signals. In addition, we need one novel parameter coming from both situations which can differentiate anisotropy and noise. Considering the normal diffusion-weighted imaging, formula (1.35) and (1.36) lead us to below equation,

$$ADC(x, y, z) = \ln[S_2(x, y, z) / S_1(x, y, z)] / (b_1 - b_2) \tag{0.67}$$

ADC is the apparent diffusion coefficient which indicates that the diffusion process is not free in tissues, but hindered and modulated by many mechanisms (restriction in closed spaces, tortuosity around obstacles, etc.). $S_2$ and $S_1$ are signals with different echo time, in our case they should be different $\tau$ . $b_1$ and $b_2$ are corresponding b factors which depend only on the acquisition parameters, in our case normally the same because only theta varies (discussed more later). So eventually we gain a heuristic conclusion that what really matters to differentiate anisotropy and noise is the natural logarithm of $S_2$ and $S_1$ ratio.

For one specific $\theta$ value which is the correlation direction, we perform two experiments of $\tau$ = 3.15 ms and $\tau$ = 4.15 ms, obtain two images of red bone marrow, calculate the natural logarithm of $S_2$ and $S_1$ ratio for each pixel in the image. So finally one image is obtained for one certain $\theta$. If we increment the $\theta$ value by 15 degree in the range of 0 degree to 180 degree, 13 images of this novel parameter can be obtained. If we select an anisotropic area in red bone marrow containing many pixels and compute the mean value among those pixels for each of those 13 images, we expect that the 13 mean values obey a reasonable curve. It should be symmetric because signals in the first half of $\theta$ should be the same as the second half of it according to $3\cos^2(\theta)-1$. For isotropic area, it is probably just random values reflecting noise.

Before showing that 13 mean values curve for a certain area containing anisotropy in red bone marrow, we would like to put all iDQC images with changing $\theta$ in line below as figure 16. It helps us to make sure our iDQC scans are correct because the signal for each basically obeys the rule $3\cos^2(\theta)-1$.

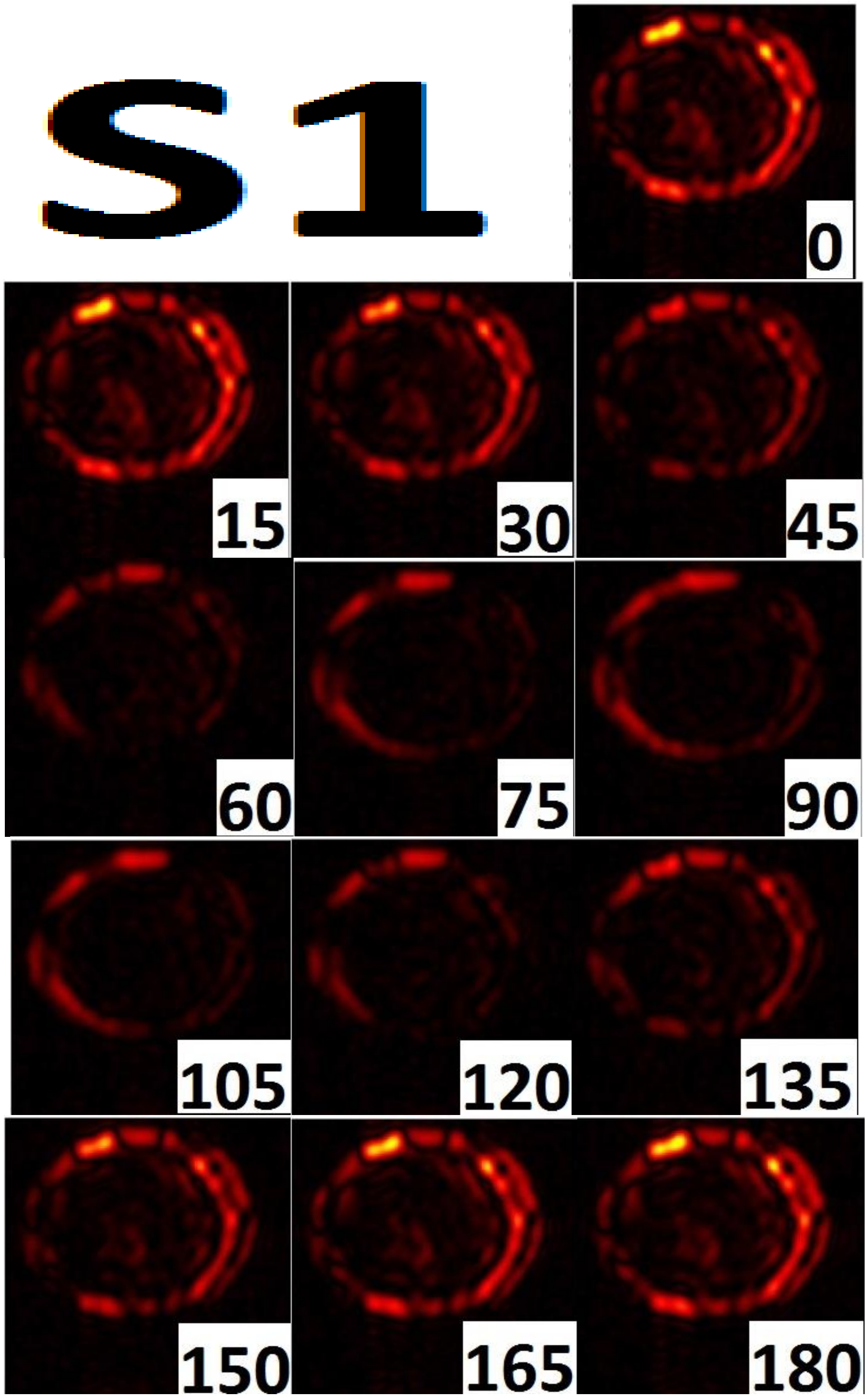

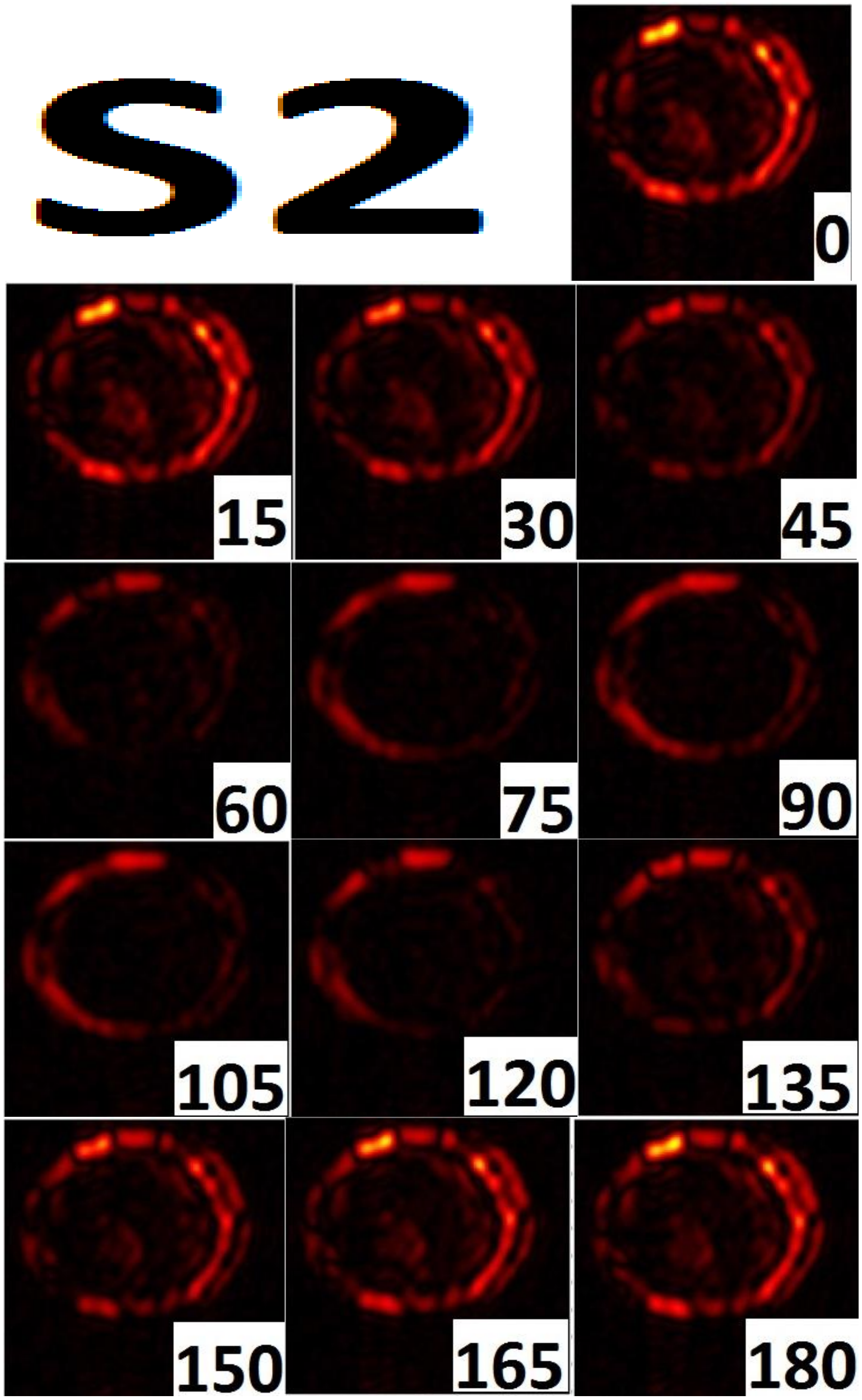

**Figure 16: For red bone marrow. The images starting with S1 are iDQC signals for $\tau$ = 4.15 ms with different correlation direction $\theta$. The images starting with S2 are iDQC signals for $\tau$ = 3.15 ms with different correlation direction $\theta$.**

In order to show the curve of those 13 values, we randomly select a small area in the central area of red bone marrow. The normalized mean of $\ln(S_1 / S_2)$ in that selected area changes with different $\theta$ value which is shown below in figure 17. It is obvious that this curve is symmetric about 90 degree. It increases from 0 degree to 90 degree and falls down from 90 degree to 180 degree. This curve is very recognizable in terms of identifying anisotropy. However, before assigning this kind of curve to anisotropy property, we need to check what this kind of curve looks like for isotropy area, as control experiments.

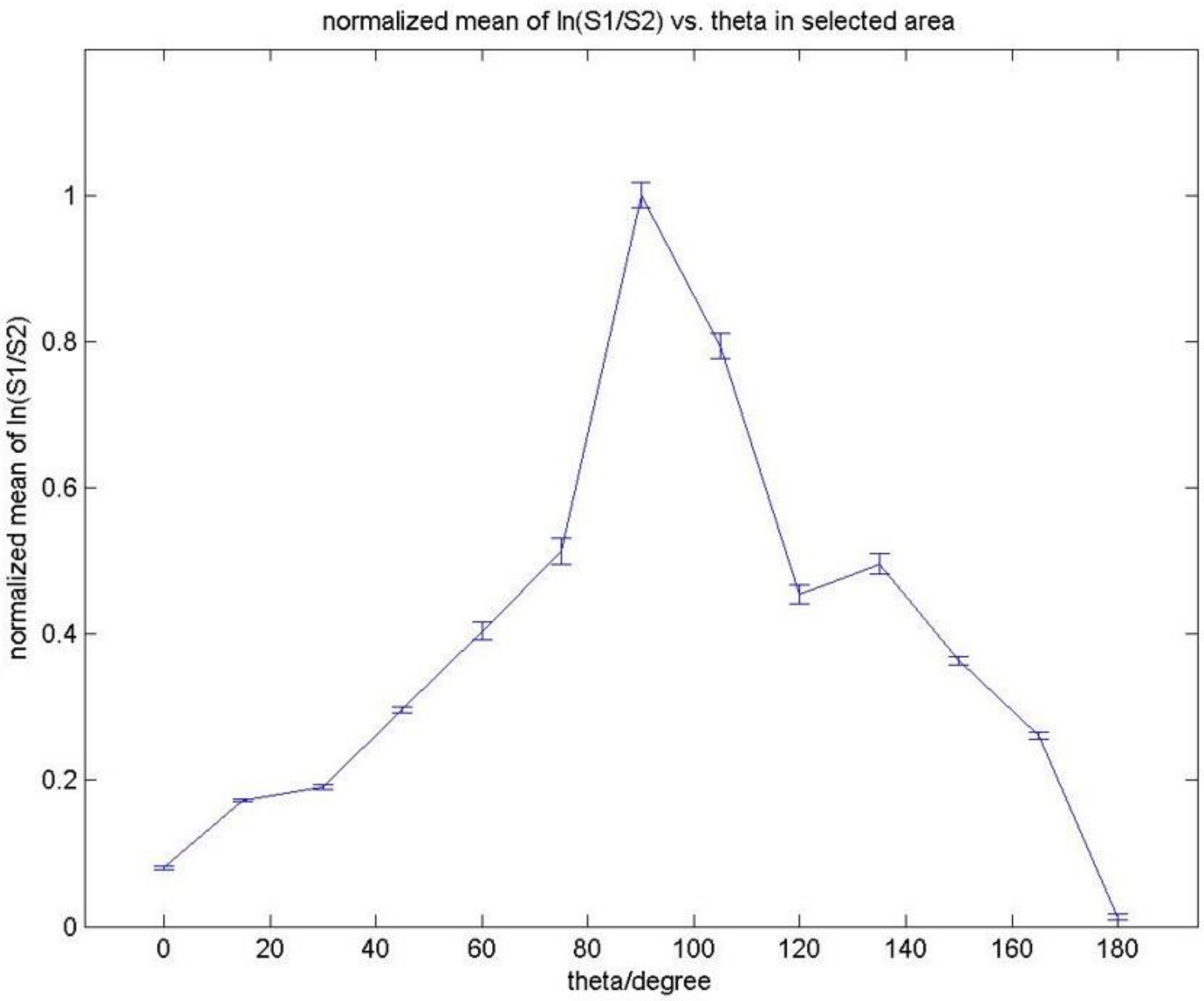

**Figure 17: We select a small area in the center of red bone marrow supposed to have anisotropy, calculate normalized mean of $\ln(S_1/S_2)$ in that area for 13 different correlation directions.**

The control experiments are performed on a tube of vegetable oil emulsion which has no anisotropy. For emulsions, noise is expected for the normalized mean of $\ln(S_1/S_2)$ on different $\theta$. First of all, all those iDQC images are shown below in figure 18. Just from the intensity, we observed that it is getting small until magical angle, then increasing until 90 degree, then decreasing until another magical angle again, then getting large until 180 degree.

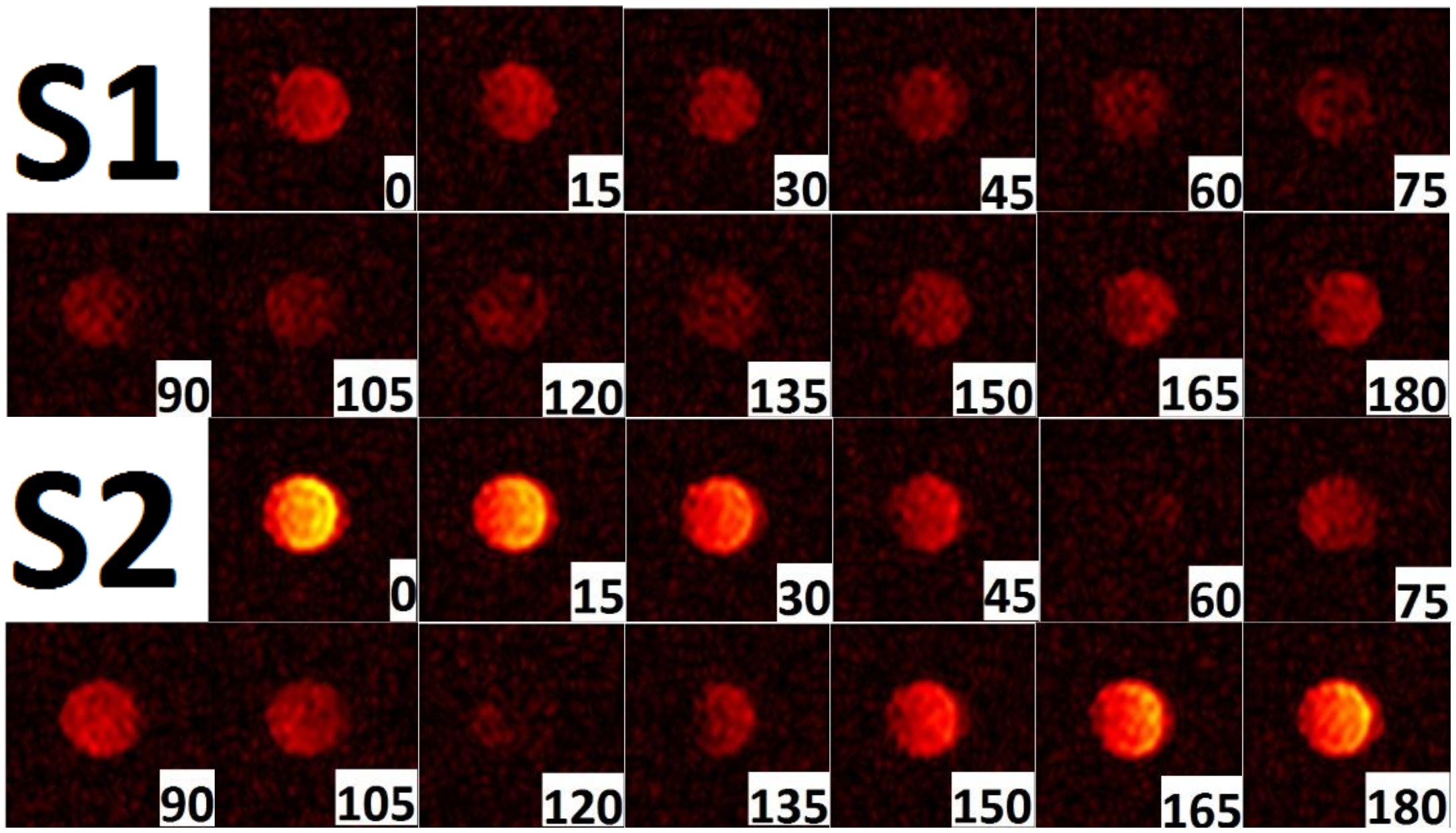

**Figure 18: For vegetable oil emulsion. The images starting with S1 are iDQC signals for $\tau$ = 4.15 ms with different correlation direction $\theta$. The images starting with S2 are iDQC signals for $\tau$ = 3.15 ms with different correlation direction $\theta$.**

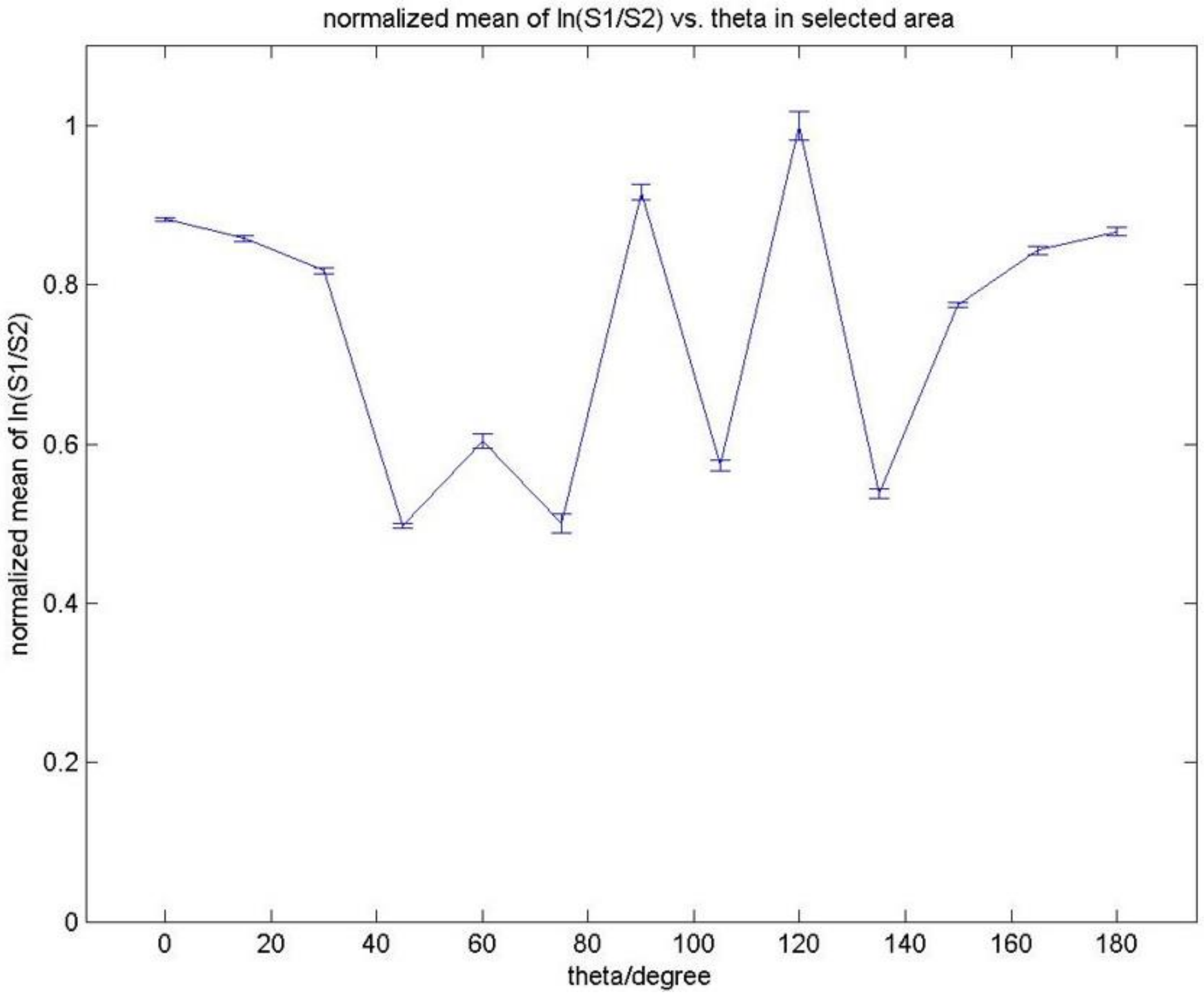

**Figure 19: Of course the vegetable oil emulsion has no anisotropy. We select a small area (the same size as Fig. 17 selected) in the center of signal, calculate normalized mean of $\ln(S_1/S_2)$ in that area for 13 different correlation directions.**

For the vegetable oil emulsion, it is easily seen that the normalized mean of $\ln(S_1/S_2)$ is just like random values with different $\theta$, and they are all almost between 0.5 and 1. So until now, it explicitly proves that anisotropic area will have a recognizable $\ln(S_1/S_2)$ curve on different $\theta$. The $S_1$ differentiates from $S_2$ by $\tau$ which is the period between 90 degree pulse and 120 degree pulse in positive two quantum iDQC.

Looking back those iDQC images, another pre-method to roughly identify anisotropy is to look at how the real component of those signals changes with $\theta$. This is shown in Fig. 20.

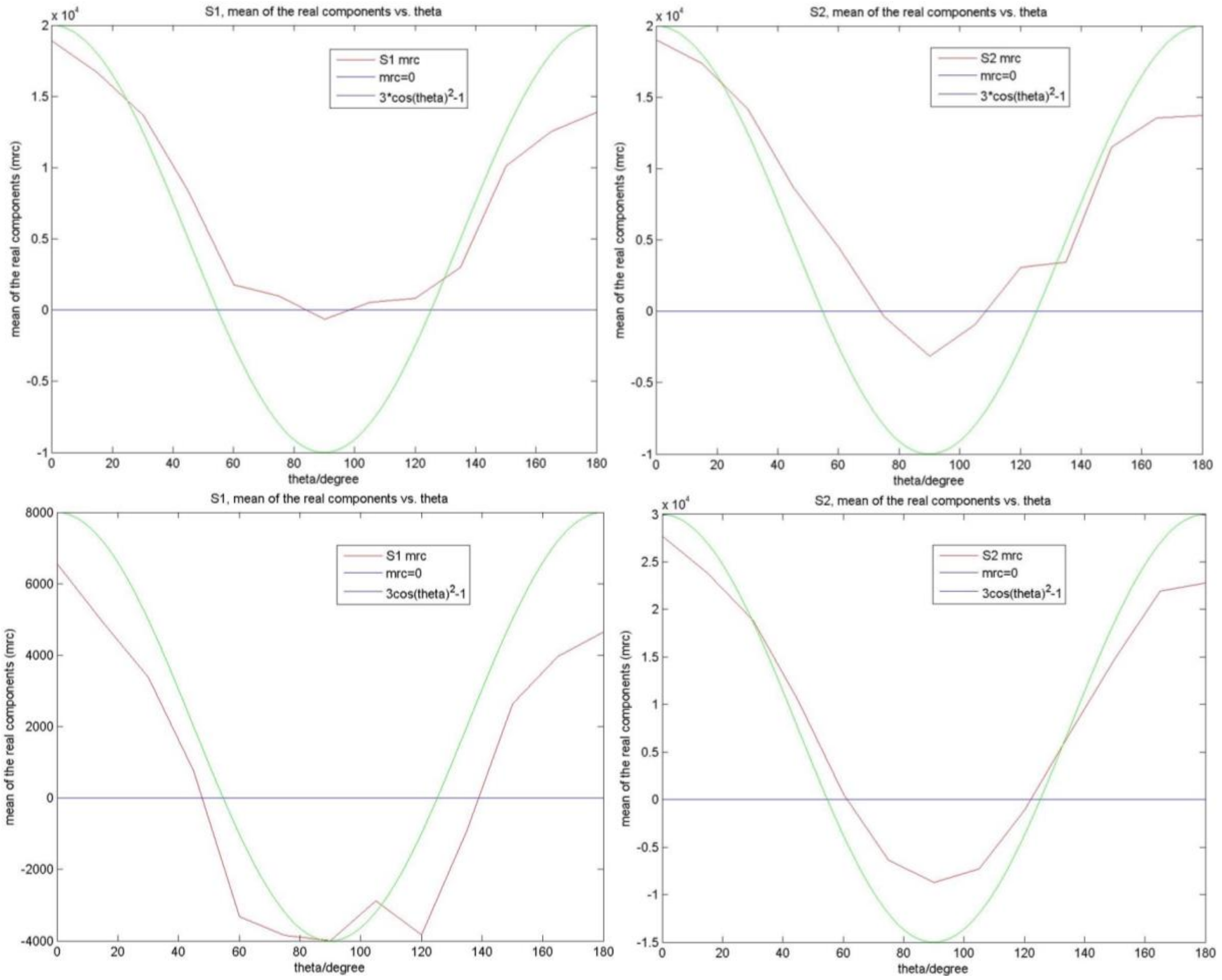

**Figure 20: The red curves: The upper left one is the real component of iDQC signal with $\tau$ = 4.15 ms in selected anisotropic area of red bone marrow, the upper right one is the same case with $\tau$ = 3.15 ms. The lower left one is the real component of iDQC signal with $\tau$ = 4.15 ms in selected isotropic area of vegetable oil emulsion, the lower right one is the same case with $\tau$ = 3.15 ms. The green curves: $c\cdot[3\cos^2(\theta)-1]$, constant c is chosen based on value at $\theta=0$ degree.**

In figure 20, the upper two plots are from anisotropic area in red bone marrow, so it is clear that they don't obey the shape of $3\cos^2(\theta)-1$, suggesting that these regions might be anisotropic. While the lower two plotting from vegetable oil emulsion fit with the shape $3\cos^2(\theta)-1$ very well. It indicates that they are pretty much associated with isotropic area, which is true.

# 5. Conclusions

In summary, the work covered in these chapters shows strong promise for the future of iMQC in biomedical MRI. We have demonstrated that the unique properties of iMQCs can be used to bring advantages in structural imaging. Future work on these projects will focus on following parts.

The first future work is to look at the effect from different correlation distance to iDQC images and corresponding anisotropy maps. It is possible that another correlation distances will better enhance the signal contrast and better show anisotropy information in the rat brain. The correlation distance can be changed by the correlation gradients strength and duration. It is highly possible that some other biological information or diseases only can be better achieved or studied by using a very narrow correlation distance. So by applying different correlation distance in mesoscopic scales, the produced iDQC images should be studied qualitatively and quantitatively.

The second future work should be to better understand the difference between Standard-Crazed and mod-Crazed images, and to create novel fractional anisotropy maps related to phase. Clearly, there will be many artifacts in those images which can be improved, such as coil phase shifts and shim effects. More importantly, they are new methods and these phase-relevant results show a total new contrast which can be co-registered with many other MR methods, and there are many applications in the future because they are looking at the proper scales 10 um to 500 um.

The third future work should be trying to apply those techniques into more clinical research. For instance, we already proved that iMQCs can show the anisotropy of the red bone marrow. The next step can be how to relate red bone marrow anisotropy with some diseases.